\documentclass[11pt]{article}

\usepackage[final]{acl}

\usepackage{times}
\usepackage{latexsym}

\usepackage[T1]{fontenc}
\usepackage[utf8]{inputenc}

\usepackage{microtype}

\usepackage{inconsolata}

\usepackage{graphicx}
\usepackage{caption} % DO NOT CHANGE THIS AND DO NOT ADD ANY OPTIONS TO IT
\usepackage{color,soul}
\usepackage{makecell}

\usepackage{times,latexsym}
\usepackage{url}
\usepackage{algorithm}
\usepackage{algorithmic}
\usepackage{booktabs}
\usepackage{subcaption}
\usepackage{amsmath}
\usepackage{framed}
\usepackage{times}  % DO NOT CHANGE THIS
\title{Computational Narrative Understanding for Expressive Text-to-Speech}

\author{
  Gaspard Michel$^{\dagger\ast}$ \\ \texttt{gmichel@deezer.com} \And Elena V. Epure$^{\dagger\diamond}$ \\\texttt{elena.epure@idiap.ch} \\ \AND Christophe Cerisara$^\ast$ \\ \texttt{christophe.cerisara@loria.fr}
  \AND
  $^\dagger$ \normalfont{Deezer Research, Paris, France} \\ $^\ast$ \normalfont{Loria, Nancy, France} \\ 
  $^\diamond$ \normalfont{IDIAP, Martigny, Switzerland}
}

\begin{document}
\maketitle

\begin{abstract}

Recent advances in text-to-speech (TTS) have been driven by large, multi-domain speech corpora, yet the expressive potential of audiobook data remains underexamined. We argue that human-narrated audiobooks, particularly fictional works, contain rich and diverse prosodic cues arising from the natural alternation between neutral narration and expressive character dialogue. Building from this observation, we introduce LibriQuote, a large-scale 5.3K hours of expressive speech drawn from character quotations.
Each quote is supplemented with contextual pseudo-labels for speech verbs and adverbs that characterize the intended delivery of direct speech (e.g., ``\textit{he whispered softly}'').
We found that fine-tuning a flow-matching model on LibriQuote yields substantial improvements in expressivity and intelligibility, while training from scratch enhances expressiveness of an autoregressive TTS model.
Benchmarking on LibriQuote-\textit{test} highlights significant variability across systems in generating expressive speech.
We publicly release the dataset, code, and evaluation resources to facilitate reproducibility.
Audio samples can be found at \url{https://libriquote.github.io/}.

% Text-to-speech (TTS) systems have recently achieved more expressive and natural speech synthesis by scaling to large speech datasets.
% However, the proportion of expressive speech in such large-scale corpora is often unclear, and existing expressive speech corpora are typically smaller in scale.
% In this paper, we introduce the LibriQuote dataset, a 5.3K hours of expressive speech drawn from character quotations derived from read English audiobooks, designed for both fine-tuning and benchmarking expressive zero-shot TTS system.
% Each utterance is supplemented with the context in which it was written, along with pseudo-labels of speech verbs and adverbs used to describe the quotation (\textit{e.g. ``he whispered softly''}).
% Training experiments show that fine-tuning a baseline TTS system on LibriQuote significantly improves speech intelligibility, while training from scratch improves expressivity.
% Besides, benchmarking with LibriQuote reveals large discrepancy in recent TTS systems ability to synthesize expressive speech, where only one model was shown to deliver highly expressive speech.
% The dataset and evaluation code are freely available.
% Audio samples can be found at \url{https://libriquote.github.io/}.

\end{abstract}

% \begin{figure*}[t!]
%     \centering
%     \includegraphics[width=.85\linewidth]{emotion2vec_test (2).png}
%     \caption{t-SNE projection of emotion vector representations computed with \texttt{emotion2vec-plus-base}. LibriQuote-\textit{test}  (a) quotations and (b) reference narration (non-quotation) utterances; (c) Subsample of LibriHeavy segments containing one quotation ($N=5734$).} % l'explication des "reference narration" ne vient que tard dans le papier et c'est donc bien de donner un hint dans la lédende sur ce que ca veut dire
%     \label{fig:emo}
% \end{figure*}

\section{Introduction}
Recently, text-to-speech (TTS) systems have significantly improved by scaling to larger, multi-domain speech corpora \cite{shen2023naturalspeech, wang2023neural, kharitonov2023speak, megatts2, anastassiou2024seed, maskgct, f5tts, sparktts}, exhibiting remarkable naturalness, quality and voice following abilities.
Typically, these large-scale corpora contain speech derived from \textit{audiobooks} \cite{librilight, pratap2020mls, libriheavy} or in-the-wild data \cite{he2024emilia}.
While the latter may include diverse expressivity and speaking styles, it is sometimes claimed that audiobooks may lack such expressiveness \cite{he2024emilia}.

Contrary to this, we argue that human readings of audiobooks, especially fictional works, offer extensive prosodic variability through a natural alternation between neutral speech and expressive utterances.
Neutral speech usually corresponds to narrative parts, the fragments enunciated by the narrator, while expressive speech can be employed during dialogues between fictional characters.
Dialogue is an essential narrative function \cite{genette1983narrative} that allows fictional characters to portray themselves and express aspects of their personality, emotional states, and world representation.
As the story unfolds, characters navigate dynamically through their own emotion arc \cite{vishnubhotla-etal-2024-emotion}, which can be portrayed during audiobook reading.
Indeed, numerous blog posts written by professional narrators emphasize the importance of \textit{impersonation}, or the act of bringing life to fictional characters \cite{gonzalesaudiobooks, brownaudiobooks, goodwinaudiobooks}.

%In this work, we put fictional characters at the forefront, and propose LibriQuote, a large-scale dataset of expressive utterances derived from LibriVox\footnote{\url{https://librivox.org}} recordings.
In this work, we put fictional characters at the forefront, by creating \textit{a large-scale dataset of expressive quotes} and \textit{proving its effectiveness for expressive TTS}.
Our dataset, \textbf{LibriQuote}, differs from standard large-scale audiobook datasets \cite{pratap2020mls, librilight, libriheavy} by relying on a narrative-aware segmentation focused on character quotes that exhibit expressivity and diversity, as opposed to narration, which corresponds to a more neutral reading style.
%by segmenting speech utterances in two ways: character utterances that involve expressive and diverse speech, and narration parts that correspond to a neutral, formal reading style.
%In contrast to prior works using LibriVox, 
%We also provide as side information the context in which utterances appears, supplemented with pseudo-labels of speech verbs and adverbs employed in the book to describe the utterance (\textit{e.g.} ``he whispered softly''), automatically annotated and further validated by humans.
We automatically annotate, and validate with humans, speech cues specific to quotations as contextual information: pseudo-labels of \textit{speech verbs} and \textit{adverbs} used in the narrative to characterize direct speech utterances (\textit{e.g.} ``he whispered softly'').
LibriQuote-\textit{train} includes 3300 speakers and 5.3K hours of quotations and 12.7K hours of narration, 
while LibriQuote-\textit{test} provides 7.4 hours of quotations from 15 speakers.
%, automatically annotated and further validated by humans.
%These pseudo-labels provide rich information that we further exploit to construct an expressive data subset.
% 
% The former typically involve expressive and diverse speech, while the latter correspond to formal reading style.
% Through an extensive alignment pipeline involving audio preparation and transcription from LibriVox, text matching from Project Gutenberg\footnote{\url{https://www.gutenberg.org/}} and quotation detection \cite{bamman2014bayesian}, we collected more than 3 million quotations uttered by fictional characters that we systematically aligned with their book counterpart, resulting in nearly 5400 hours of speech.

% An overview of the pipeline is provided in Figure~\ref{fig:intro}.
% , providing as side information the context in which they appear, supplemented with pseudo-labels of the speech verb and adverb used to describe the utterance (\textit{e.g.} ``he whispered softly'') automatically annotated and further validated by humans.
% Qualitative analysis of the data reveals that LibriQuote contains a large set of emotions and accents, facilitating the training and benchmarking of expressive TTS systems, as depicted in Figure~\ref{fig:emo}.
Our findings are the following:
\begin{enumerate}
    \item \textit{Fine-tuning} a state-of-the-art autoregressive TTS system with LibriQuote-\textit{train} does not directly improves expressivity, but yields \textit{substantial improvements in speech intelligibility} on in-domain and out-of-domain data (Section~\ref{sec:training}). Besides, when \textit{training from scratch} the same TTS system  with LibriQuote-\textit{train}, we \textit{increase speech expressivity}, but at the cost of reduced intelligibility due to substantially reduced training data \cite{ye2025llasascalingtraintimeinferencetime}.
    In contrast, \textit{fine-tuning a flow-matching model directly enhances both speech expressivity and intelligibility}.
    Finally, we show that \textit{using both narrations and quotations} contained in LibriQuote-\textit{train} is \textit{a promising direction for training expressive and intelligible TTS systems}.
    % These experiments indicate that integrating our training dataset with additional datasets in a curriculum-learning framework could be a suitable strategy to increase expressivity, while maintaining a high intelligibility.
    \item The benchmarking of various TTS systems on LibriQuote-\textit{test} reveals large discrepancies in TTS systems’ ability to synthesize expressive speech (Section~\ref{sec:benchmark}). This indicates that LibriQuote-\textit{test} represents a challenging benchmark that could foster advances in developing expressive TTS.
\end{enumerate}

The dataset is available at the following links:

\noindent  Huggingface: {\small \url{https://huggingface.co/datasets/gasmichel/LibriQuote}}

\noindent  GitHub: {\small \url{https://github.com/deezer/libriquote}}
% Experiments show that using LibriQuote-\textit{train} for fine-tuning an existing state-of-the-art autoregressive TTS system yields clear improvements in speech intelligibility on out-of-domain data, while training from scratch with LibriQuote-\textit{train} improves speech expressivity on character quotations.
% Moreover, quantitative and human evaluation when benchmarking on LibriQuote-\textit{test} reveals large discrepancies in TTS systems’ ability to synthesize expressive speech.

%We release publicly the dataset, code and evaluation samples used in objective and subjective experiments.
% to encourage further benchmarking and development of more expressive TTS systems.  

\section{Background and Motivation}

\subsection{Related Speech Datasets}
Audiobooks have long been the de-facto open-source data for training TTS systems.
The LibriVox project records thousands of public-domain books read by volunteer speakers.
LibriSpeech \cite{librispeech}, a 1000-hour English corpus covering 2400 speakers and extracted from LibriVox, has been widely used to train and benchmark TTS systems.
LibriTTS \cite{libritts} offers an enhanced version of LibriSpeech with better audio quality and utterance segmentation, though with less training data, while LibriSpeech-PC \cite{librispeechpc} restores punctuation and capitalization in LibriSpeech transcripts.

Libri-Light \cite{librilight} is the largest open English audiobook corpus at about 60,000 hours of LibriVox speech but lacks audio transcriptions.
The MLS dataset \cite{pratap2020mls} includes 50,000 hours of multilingual audiobooks and provides transcriptions aligned with the original text.
LibriHeavy \cite{libriheavy} improves Libri-Light by aligning audio segments with their corresponding book text, but does not provide contextual book information around segments.
These 30-second segments are cut at sentence boundaries and may include multiple quotations and narration, alternating between neutral (narration) and expressive (quotation) speech.

Moreover, several efforts towards the creation of expressive read speech corpora have been made. Emotional speech datasets \cite{iemocap, cremad, ravdess, esd} that assign discrete emotion labels to speech utterances have been extensively used to evaluate emotional synthesize of TTS systems \cite{megatts2, maskgct, sparktts}.
The L2-ARCTIC corpus \cite{l2arctic} proposes an 11 hour corpus with ten speakers, spanning 5 different accents.
EXPRESSO \cite{expresso} introduces an high-quality dataset of read and improvised speech across 26 expressive styles.

\subsection{Gaps in Current Datasets}
\begin{figure}
    \centering
    \includegraphics[width=0.85\linewidth]{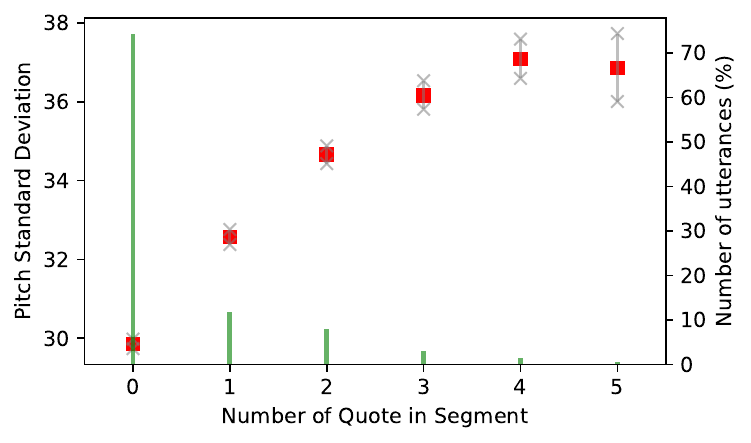}
    \caption{Average pitch standard deviations (red squares) per number of quotations in audio segments in the LibriHeavy-\textit{small}, with bootstrapped 95\% confidence intervals (grey lines); and percentage of total number of segments per number of quotations (green).}
    \label{fig:pitch_std}
\end{figure}
Prior works that use audiobooks as a resource for speech data, such as LibriTTS or LibriHeavy, have overlooked aspects of narrative discourse.
They either completely discard character quotations, or combine them with other quotations and/or neutral narrative parts in their audio segments, which are constructed from random sentence breaks.
% Instead, these works have explored segmenting audio utterances at sentence breaks.
% We argue that this semi-random segmentation is prone to include a combination of narrative parts and possibly multiple quotations uttered by different characters.
Such segments are thus likely to exhibit multiple "distinct" pitch distributions, possibly alternating between neutral, emotional and/or pitch-varying speech.
We illustrate this effect in Figure~\ref{fig:pitch_std}.

Specifically, we performed quotation detection on 120000 segments from LibriHeavy-\textit{small}, and calculated their pitch standard deviation in the corresponding audio segments.
We found that 75\% of segments are only narrative parts, while the remaining 25\% contain from 1 to 12 quotations.
We see that pitch standard deviation typically increases with the number of quotations within the segment (Spearman $\rho = 0.218$, $p<0.001$).

Modeling these complex pitch distributions is a challenging task. 
It can hinder the training of TTS systems, which may end up focusing on learning to model the simpler, narrative parts instead.
By shifting the focus from formal narration to character quotations, LibriQuote provides varied direct speech samples that span a broader range of emotions, at scale.
These traits are desirable for generating high-quality, human-like speech \cite{he2024emilia}.
Moreover, we supplement these quotations with narrative information indicating \textit{how} an utterance should be spoken, providing a natural language pseudo-label for expressivity.

\section{The LibriQuote Dataset}

%This section describes the steps involved in preparing LibriQuote, followed by descriptive statistics. Then we present our methodology to extract narrative information for each utterance. 
Distinguishing between "story" (what is happening) and "discourse" (how it is told) is a key aspect of narrative understanding.
NLP research has extensively focused on various aspects of \textit{narrative discourse}, as theorized by \citet{genette1983narrative}.
Recently, a variety of works explored the usage of Large Language Models (LLMs) to model elements of narrative understanding, such as dialogue \cite{michel-etal-2025-evaluating}, character roles and information extraction \cite{stammbach-etal-2022-heroes, gurung-lapata-2024-chiron}, narrativity \cite{hatzel-biemann-2024-story} or structure \cite{soni-etal-2023-grounding}.

The current work is grounded in the perspectival issues of fictional characters, including point of view and dialogues, anchored in the \textit{voice} linking function of Genette's theoretical framework.
In fiction, utterances are sometimes supplemented with \textit{narrative information} in the form of speech verbs and adverbs (\textit{e.g. ``he whispered softly''}).
These indicators have been shown to produce positive effects during reading, such as enhancing reading comprehension \cite{wolters2022reading} or character identification \cite{van2017evoking}.
For audiobook narrators, they are a valuable resource to set the tone and point of view of the character.
Other contextual information such as the emotional context of a character are likely to also impact speech delivery, but are often prone to subjective interpretation and are harder to automatically infer \cite{vishnubhotla-etal-2024-emotion}.

% Figure~\ref{fig:intro} provides an overview of the full process.

\subsection{Data Creation Pipeline}
Our data creation pipeline is similar to the one proposed for LibriHeavy \cite{libriheavy}. %to align audio files downloaded from LibriVox to their original text.
However, we filter out books that are unlikely to contain fictional characters, such as bibliographical or other types of non-fictional works.
We detail the filtering and audio-text alignment stages below.

\subsubsection{Audio Preparation}
We start by browsing the LibriVox API\footnote{\url{https://librivox.org/api/info}} to collect recording information.
We only extract English recordings corresponding to Fiction books.
This step is necessary to filter out non-fiction books that are unlikely to contain fictional characters.
Each recording is downsampled to 16kHz for further processing, but we provide links to download the original audio files as well, allowing LibriQuote to support systems operating on higher audio quality. LibriVox also includes \textit{Dramatic Readings}, where each fictional character is performed by a different speaker.
We exclude these recordings from LibriQuote, as they would require an additional speaker diarization step.\footnote{We applied the same alignment process to these recordings and plan to release these alignments in future work.}

%We also applied the alignment process described below to these recordings, but do not include them in LibriQuote as they would require an additional speaker diarization step. 
%We plan to release these alignments in future works. 

\subsubsection{Text Preparation}
For each audio recording, we download the corresponding book text from Project Gutenberg.
%Some books have external text providers, but we found by manual inspection that automatically
%downloading from these websites was either too complicated, or resulted in texts with very poor quality often due to OCR errors.
%In such cases, we browse Project Gutenberg to search for the target book, and discard it if not available on the platform.
Some books have external text providers, but manual inspection revealed that automatically downloading from these websites was often too complicated or yielded low-quality texts, frequently due to OCR errors. 
In such cases, we turned to Project Gutenberg to search for the target book and discarded it if it was not available on the platform.
We then perform quotation detection using BookNLP\footnote{\url{https://github.com/booknlp/booknlp/blob/main/booknlp/english/litbank_quote.py}} on each book.
We discard every book containing less than $20$ quotations, as we found that these books typically yield quotation detection errors.
For other books, we keep quotation positions in the original text for further alignments.

\begin{table}[t!]
    \centering
    \small
    \setlength{\tabcolsep}{5pt}
    \begin{tabular}{l|ccc|c|c}
    \toprule
         & \multicolumn{3}{c|}{\textbf{Train}} & \textbf{Dev}  & \textbf{Test}  \\
         \midrule
         & $\mathbf{N}$ & $\mathbf{Q}$ & $\mathbf{Q}_f$ & $\mathbf{Q}$ & $\mathbf{Q}$ \\
         \midrule
    Count & $3,87$M & $3,51$M & $378$K & $2921$  &  $5598$\\
    Hours & $12723$ & $5359$ & $379$ & $5.1$  & $7.4$  \\
    Dur (s) & $11.8$ & $5.5$ & $3.6$ & $6.2$ & $4.8$\\
    \midrule
    Speakers & \multicolumn{2}{c}{$3314$}  & $2818$ & $56$ & $15$\\
    $\;\;$ avg (h) & $3.8$ & $1.6$ & $0.13$ & $0.1$ & $0.5$\\
    % $\;\;$ avg (\#) & $3.8$ & $1.6$ & $0.13$ & $0.1$ & $0.5$\\

    \midrule
    Books & \multicolumn{2}{c}{$2991$} & $2900$ & $65$ & $27$\\
    \bottomrule
    
    \end{tabular}
    \caption{Descriptive statistics of LibriQuote. Count  is the utterance number. $\mathbf{N}$ refers to narration, $\mathbf{Q}$ to quotations and $\mathbf{Q}_f$ to filtered quotation ($\mathbf{Q}_f\subset\mathbf{Q}$).}
    \label{tab:desc_stats}
\end{table}
\subsubsection{Audio Transcription}
% In order to align audio segments to their corresponding text segments, we carry out audio transcription.
% following LibriHeavy \cite{libriheavy}.
We start by segmenting each audio recording into 30-second segments, with a 2 second overlap at each side.
These segments are then transcribed with a Zipformer-Transducer ASR model trained on LibriSpeech\footnote{https://github.com/k2-fsa/icefall/pull/1058}, and combined by leveraging word-level timestamps to obtain a full timestamped transcript of the recording. 

\subsubsection{Text-Audio Alignment}
% We largely follow LibriHeavy to perform the alignment between the transcript-book pairs.
Since audio recordings often involve a full chapter, this step is necessary to align the corresponding chapter text within the full original book.
% and to have fine-grained alignment at the word level.
This alignment process involves two stages.
In the first stage, \textit{close matches} between a transcription and the book text are constructed.
These \textit{close matches} correspond to $(i,j)$ pairs where $i$ is an index in the transcription and $j$ an index in the book text.
Then, a coarse alignment is produced by taking the longest chain of pairs $(i_1, j_1),
\dots,(i_N,j_N),$ such that $i_1  \leq \dots  \leq i_N$ and $j_1  \leq \dots  \leq j_N$.
The second stage produces a final alignment by concatenating Levenshtein alignments \cite{lcvenshtcin1966binary} between the transcription and the text segment produced by the longest chain of pairs.
We refer the reader to \citet{libriheavy} for additional information regarding the alignment process.

% \begin{table}[t!]
%     \centering
%     \small
%     \begin{tabular}{l|cc|cc|cc}
%     \toprule
%          & \multicolumn{2}{c}{\textbf{Train}} & \multicolumn{2}{c}{\textbf{Dev}}  & \multicolumn{2}{c}{\textbf{Test}}  \\
%          \midrule
%          & \textbf{N} & \textbf{Q} & \textbf{N} & \textbf{Q} & \textbf{N} & \textbf{Q} \\
%          \midrule
%     Count & $3.9$M & $3.5$M & $4065$ & $2921$  & $5803$ & $5803$\\
%     Hours & $12723$ & $5359$ & $15.3$ & $5.1$  & $10.8$ &$7.5$  \\
%     % Dur (s) & $11.8$ & $5.5$  &  $13.5$ & $6.2$ & $10.1$ & $5.8$\\
%     \midrule
%     Speakers & \multicolumn{2}{c|}{$3314$}  & \multicolumn{2}{c|}{$56$} & \multicolumn{2}{c}{$15$ (8m, 7f)}\\
%     $\;\;$ avg (h) & $3.8$ & $1.6$ & $0.3$ & $0.1$ & $0.72$ & $0.5$\\
%     \midrule
%     Books & \multicolumn{2}{c|}{$2991$} & \multicolumn{2}{c|}{$65$}  & \multicolumn{2}{c}{$27$}\\
%     \bottomrule
    
%     \end{tabular}
%     \caption{Descriptive statistics of LibriQuote. \textbf{N} refers to narration and \textbf{Q} to quotations. In this table, $1$M corresponds to one million. The test set contains 8 male speakers (m) and 7 female speakers (m). }
%     \label{tab:desc_stats}
% \end{table}

\subsubsection{Quotation Alignment}
The output of the text-audio alignment is an array of words in the original text, timestamped in the audio recording.
We leverage the timestamps and the output of the quotation detection system to segment each audio into a set of read quotations.
Additionally, we build audio segments from narration paragraphs (\textit{i.e.} parts that are not quotations) in each book, and trimmed long start end silences for each audio segment.
The final construction stage involved assigning contextual information around quotations.
Based on the alignments, we match each quotation with a window of around $100$ words occurring before and after it, with boundaries defined by paragraphs.
With LibriQuote, we release the preprocessed texts as well as full paragraph and quotation alignments.
Thus, it is possible to derive a contextual window of any desirable length around quotations.
% Note that in this case, a single quotation is seen as a unique paragraph (\textit{i.e.} we do not split the contextual information within quotations or paragraphs).

\subsection{Dataset Characteristics}
\label{sec:desc_stats}
% \begin{figure}
%     \centering
%     \includegraphics[width=\linewidth]{train_hist-5.pdf}
%     \caption{Utterance duration (s) of LibriQuote-\textit{train}.}
%     \label{fig:enter-label}
% \end{figure}

Table~\ref{tab:desc_stats} shows LibriQuote's descriptive statistics.

\paragraph{Training Set}
The filtering and alignment process resulted in a total of 12723 training hours of narration and 5359 training hours of quotation across 2991 unique books and 3314 unique speakers.
As expected, narration parts constitute the majority of audiobook recording in terms of duration, but quotations still represent almost 50\% of the total number of training utterances.
Indeed, the average audio segment duration is 11.8 seconds for narration parts and 5.5 seconds for quotations.
% Figure~\ref{fig:enter-label} displays the histogram of utterance durations.
LibriVox does not provide gender information of its readers, so the balance between male and female speaker in the training set is unclear.

% \begin{table}[t!]
%     \centering
%     \small
%     \begin{tabular}{l|ccc|ccc}
%     \toprule
%     & \multicolumn{3}{c|}{\textit{Verbs}} & \multicolumn{3}{c}{\textit{Adverbs}} \\
%     \midrule
%     Support & \multicolumn{3}{c|}{344} & \multicolumn{3}{c}{56} \\
%     \midrule
%     & \textbf{P} & \textbf{R} & \textbf{F} & \textbf{P} & \textbf{R} & \textbf{F} \\
%     \midrule
%         Llama3.3-70b & 0.82 & 0.97 & 0.89 & 0.65 & 0.86 & 0.74 \\
%          \;\;$+ \textit{Conf}$ & 0.91 & 0.91 & \textbf{0.91} & 0.8 & 0.54 & 0.65 \\
%         \midrule
%         Qwen2.5-72b & 0.88 & 0.88 & 0.88 & 0.62 & 0.85 & 0.71 \\
%         \;\;$+ \textit{Conf}$ & \textbf{0.98} & 0.8 & 0.88 & \underline{0.93} & 0.73 & \underline{0.82} \\
%         \midrule
%     Phi-4 & 0.85 & 0.93 & 0.89 & 0.74 & 0.81 & 0.77  \\
%     \;\;$+ \textit{Conf}$ & \textbf{0.92} & 0.86 & 0.89 & \textbf{0.95} & 0.61 & 0.74\\
%     \bottomrule
%     \end{tabular}
%     \caption{Narrative information extraction results with Phi-4, and after filtering by self-reported confidence (Phi-4$_\text{conf}$).\textbf{P}, \textbf{R} and \textbf{F} refers to precision, recall and f1-score.}% je pense qu'il faut definir Phi4_conf pour comprendre sans chercher dans tout le texte
%     \label{tab:prompt1}
% \end{table}

\paragraph{Dev \& Test Sets}
% We designed the dev and test sets to reflect the following challenge in expressive TTS: \textit{given a reference timbre from neutral speech and the text to generate, can we synthesize a quotation with the correct level of expressiveness?}
We designed the dev and test sets to evaluate whether synthesized speech in quotations have the correct level of expressiveness.
Each quotation serves as the expressive speech, and we match them with the nearest narration utterance from the book context that acts as the neutral reference speech.
% The narration utterance acts as the neutral reference speech while the quotation acts as the expressive speech to synthesize.
Synthesizing these quotations can be particularly challenging, as the text alone may be misleading with regard to the true emotion, which is often specified by the surrounding context.
%Without context, TTS for these quotations could be particularly hard as only their text
%may be misleading with regard to the true emotion, which may rather be specified in the context.
As an example, the exclamation mark in the quotation \textit{``It's so good!'' he whispered softly} might suggest it shall be uttered with a loud voice, but the narrative information clarifies that it is actually being whispered.
Thus, we also provide contextual information in the vicinity of quotations to encourage further research on predicting expressivity from textual cues for audiobook TTS.
The test set contains 8 male and 7 female unseen speakers, for a total of 7.5 hours.

The dev set contains random utterances from speakers and books in the train set.
It thus does not encompass zero-shot speech synthesis, but aims at evaluating expressive synthesis from neutral speech for seen speakers.
However, we ensure no speaker-overlap with the \textit{test} set.

\begin{figure}
    \centering
    \includegraphics[width=\linewidth]{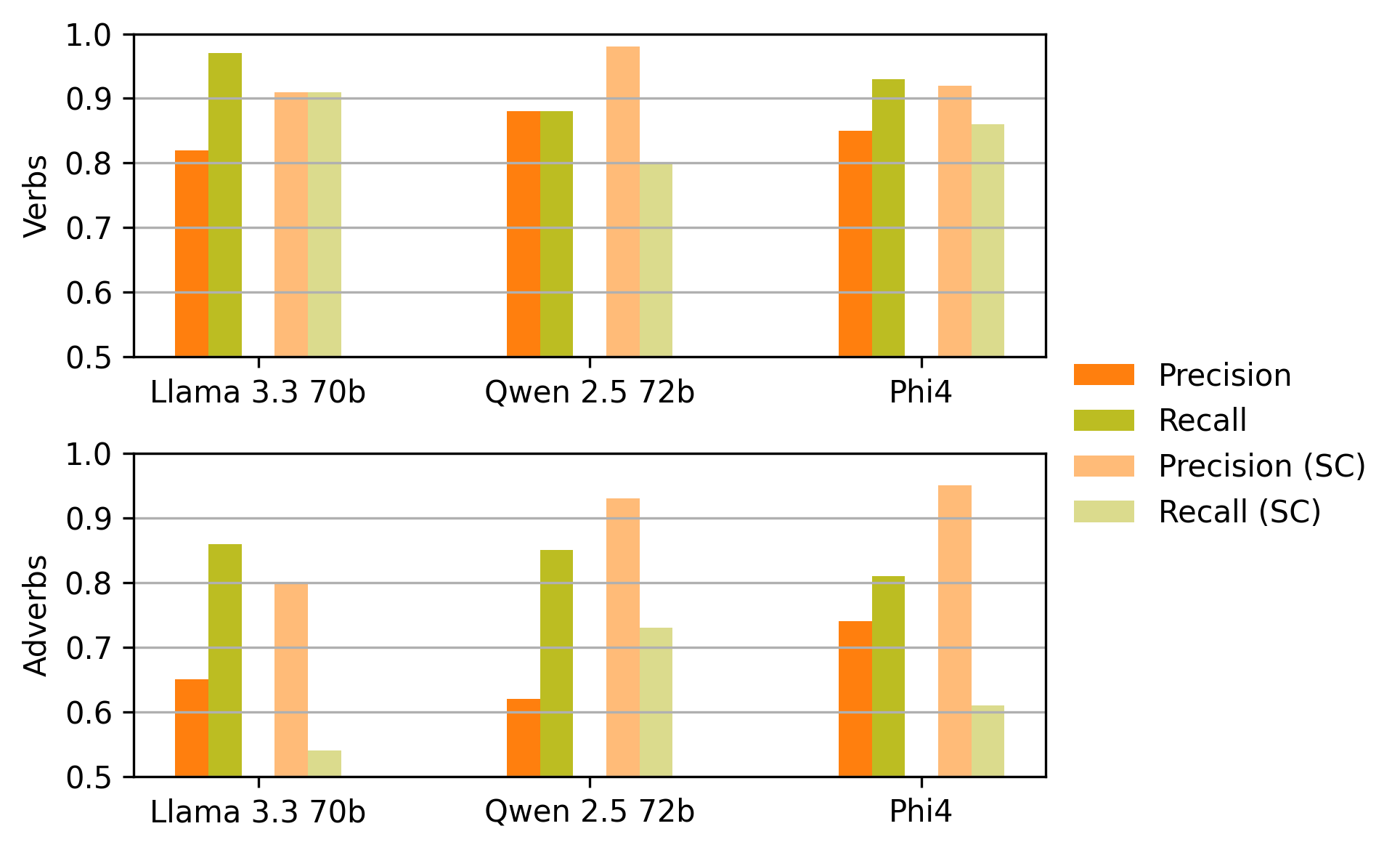}
    \caption{Evaluation of the extraction of Verbs (top) and Adverbs (bottom). 
    SC indicates self-confidence.}
    % LLMs were tested with a normal and a self-reporting confidence prompt (indicated as \textit{SC)}.}
    \label{fig:placeholder}
\end{figure}

\paragraph{Test Set Analysis}

% \hl{Tell that we remove 205 utterances with < 2 words (to remove examples such as "Hey, Jean"), and say that we analyzed manually examples with high WER (typically very prononced accent, with accent written in text: e.g. ("ain’ dat Misto Tom?") and low UTMOS (typically very pronunced emotion intensity such as sadness or anger) . Give UTMOS average value of 3.48 for GT. thus we decidec to keep these examples in the benchmark}

We start the exploration of LibriQuote-\textit{test} by computing Word Error Rate (WER) whith \texttt{Whisper-Large v3} and an automated metric for speech naturalness, UTMOS \cite{saeki2022utmosutokyosarulabvoicemoschallenge}, ranging from 1 (unnatural) to 5 (very natural).
We found that utterances exhibiting high WER often contain only a few words and proper names such as ``\textit{Hey, Jean}''.
We thus decided to remove 205 utterances from the test set that contain less than two words.
Other segments with high WER were sometimes spoken with a strong accent, which might affect Whisper transcription, but we decided to keep these segments.
%Other segments with high WER were sometimes delivered with a strong accent, which might throw off Whisper transcription, but decided to keep such segments.
UTMOS analysis reveal an average value of 3.48, where most lower valued utterances occur when a quotation is spoken with very high emotion intensity.
We thus chose not to filer out these utterances, as they correspond to desired expressivity.

To further analyse LibriQuote test set, we explore which emotions are conveyed in quotations and reference narrations.
Following \citet{maskgct}, we compute emotion representations using Emotion2Vec\footnote{\url{https://huggingface.co/emotion2vec/emotion2vec_plus_base}} \cite{ma-etal-2024-emotion2vec}.
% and accent representations using an ECAPA-TDNN model trained on Common Voice\footnote{\url{https://huggingface.co/Jzuluaga/accent-id-commonaccent_ecapa}} \cite{ ardila-etal-2020-common, zuluaga2023commonaccent}.
% We display t-SNE projections of emotion representations for LibriQuote-\textit{test} quotations and their reference narration utterance in Figure~\ref{fig:emo}.
% For comparison, we also display t-SNE projections of 5734 utterances randomly drawn from all LibriHeavy segments that contain at least one quotation.
We found that LibriQuote-\textit{test} quotations convey a larger set of predicted emotions compared to their narration reference and LibriHeavy segments: only 67\% of quotations are predicted as neutral against 87\% and 91\% for the reference and LibriHeavy respectively.
% Besides, we found that accent representations are well clustered, indicating a broad coverage of English accents.
Further details can be found in Appendix~\ref{app:emo}.

% Thus, by extracting these narrative cues and filtering out less expressive utterances (\textit{e.g. ``she said''}), we can derive a high-quality subset that spans a broad prosodic range.
% Below, we describe our method to extract these narrative information.

\subsection{Contextual Cues for Direct Speech}
\label{sec:narrinfo}

We extract narrative cues indicating the speaking style of a quotation (described by the narrator), using a contextual window spanning one paragraph before and after the quotation.
For this, we rely on several LLMs as they have shown promising results in literary dialogue understanding \cite{piper-bagga-2024-using, michel-etal-2025-evaluating}.
%Our experiments involved prompting several LLMs to 

We use few-shot prompting with 5 examples.
% Prompt details can be found in our repository.
To mitigate extraction errors and keep only informative narrative content, we replace all quotations in context with special markers, such that only narration and structure remains.
Our experiments involved prompting to self-report a \textit{confidence score} on a 1-to-10 scale that we used to prune-out unconfident predictions in order to maximize precision.
More details are provided in Appendix~\ref{app:narr_info}.

% (Phi-4$_\text{conf}$).

\paragraph{Validation} 
In an initial phase dedicated to developing guidelines and solving complex cases, two annotators independently tagged 100 quotations. 
After discussion and once satisfactory agreement was reached (we report a Cohen's $\kappa$ score of 0.87 for verbs and adverbs), a single annotator continued with 300 extra instances, yielding a total of 400 quotations annotated with \textit{verbs}, \textit{adverbs}, \textit{nouns}, and \textit{adjectives}, drawn from the train split.
 We found that adjectives and nouns were relatively rare in the annotated data, and that LLMs failed entirely to extract them. We thus report results only for verbs and adverbs in Figure~\ref{fig:placeholder}.

%In an initial phase focused on the development of guidelines and discussing complex cases, two annotators tagged independently 100 quotations.
%After discussions and once a satisfactory  agreement was reached, a single annotator continued 300 extra instances, resulting in 400 quotations annotated with \textit{verbs}, \textit{adverbs}, \textit{nouns} and \textit{adjectives}.
% Results for verbs and adverbs are displayed in Figure~\ref{fig:placeholder}.
%We found that the amount of adjectives and nouns was relatively low in the annotated data, and that LLMs completely fail at extracting these, and thus report only performance for verbs and adverbs in Figure~\ref{fig:placeholder}.
% Results for other LLMs, nouns and adjectives and runtimes can be found in the Technical Appendix.

We see that most models perform relatively well at extracting narrative cues, but still suffers from hallucinations, in particular for adverbs.
When pruning out all predictions with a confidence score lower than 10, all models exhibit higher precision at the cost of lower recall for both verbs and adverbs.
A high precision is desirable: we want to avoid quotations to be tagged with wrong expressive speech verbs and adverbs.
We thus select Phi-4 as the base LLM for further extraction of contextual cues on the full corpora, as it obtained the highest precision score on adverbs and a satisfactory precision on verbs, while also being the smallest and fastest model.

\paragraph{High-Expressivity Split}
After extracting contextual cues on the full dataset with the strategy described above, we build a high-expressivity subset.
%by selecting quotations described with meaningful cues.
First, we include all quotations that had a non-empty adverb pseudo-label regardless of the predicted speech verb, as adverbs often give precise information on how a 
quotation is uttered (\textit{e.g.} ``\textit{he said cautiously}'').
% We extract contextual cues on LibriQuote-\textit{train} using Phi-4 with Self-Confidence, yielding a (potentially empty) set of predicted verbs and adverbs for each quotation.
% To produce a filtered subset $\mathbf{Q}_f$, we first include all quotations that had a non-empty adverb pseudo-label regardless of the predicted speech verb, as adverbs often give precise information of how a quotation is uttered (\textit{e.g.} ``he said cautiously'').
Then, based on a manually defined set of \textit{expressive} speech verbs\footnote{The full list of verbs is available in Appendix~\ref{app:filtering}.}, we add all quotations that have a verb falling into that list.
% filter out all quotations () that have no predicted verbs or that have a predicted verb that did not fall into that list.
% We also include all quotations that had an adverb pseudo-label regardless of the predicted speech verb, as adverbs often give precise information of how a quotation is uttered (\textit{e.g.} ``he said cautiously'').
The resulting split, $\mathbf{Q}_f$, contains 377,776 quotations (11\% of the full quotation train set) for a total of 379 hours,
which may be used for data-efficient expressive TTS, or can foster automatic analysis of the alignment between written narrative cues and uttered speech.

\begin{table}[t!]
\centering
\small
\setlength{\tabcolsep}{3pt}
\renewcommand{\arraystretch}{1.1}
\begin{tabular}{l|ccc|cc}
\toprule
 & \textbf{WER} & \textbf{SIM-O}  & \textbf{E-Sim} & \textbf{CtxMOS}  & \textbf{WR} \\
\midrule
GT & 6.5 & - & - & 3.55$\phantom{^\star\star}$  & - \\
\midrule
\midrule
SparkTTS & 4.8$\phantom{^\star}$ & 0.46$\phantom{^\star}$ &  0.69 & 2.94$\phantom{^\star\star}$ & 38 \% \\
\midrule
%\\
% \midrule
% \midrule
% \midrule
    FT ($\mathbf{Q}_f$)  & \textbf{4.6}$\phantom{^\star}$ & 0.47$^\star$ & 0.71 & 2.97$\phantom{^\star\star}$ & 41\%\\
    FT ($\mathbf{Q}$)  & 5.9$^\star$   & 0.46$^\star$ & 0.71 & 2.89$\phantom{^\star\star}$ & 37\%\\
    Scratch ($\mathbf{Q}$) & 9.5$^\star$  & 0.40$^\star$ & 0.71 & 3.09$^{\star\star}$ & 38\% \\
    $\;\;+$ Ctxt  & 8.2$^\star$  & 0.40$^\star$ & 0.71 & 3.15$^{\star\phantom{\star}}$ & 35\%\\
    Full ($\mathbf{N} \cup \mathbf{Q}$) & 5.1$^{\star}$ & 0.41$^\star$ & 0.71 & 3.30$^{\star\phantom{\star}}$  & 37\% \\
    \;\; + FT ($\mathbf{Q}_f$) & 5.1$^{\star}$ & 0.41$^\star$ & 0.71 & 3.30$^{\star\phantom{\star}}$  & \textbf{41\%} \\
\midrule
\midrule
F5-TTS & 6.9$\phantom{^\star}$ & 0.53$\phantom{^\star}$ &  0.71 & 2.95$\phantom{^\star}$ & 31 \% \\
\midrule
FT ($\mathbf{Q}_f$)  & 6.6$^\star$ & \textbf{0.54}$^\star$ &  0.71 & \textbf{3.33}$^\star$ & 26 \% \\
\bottomrule
\end{tabular}
\caption{Evaluation on LibriQuote-\textit{test} set of various training setups. Two-sided paired student $t$-test are conducted against SparkTTS or F5-TTS ($^\star$ indicates $p<0.05$, and $^{\star\star}$ indicates $p<0.1$).}
% FT means Fine-Tuned and CtxMOS and WR indicates ContextMOS and Win-Rate respectively, while Scratch and +Ctxt indicates SparkTTS trained from scratch with and without context.}
\label{tab:res2}
\end{table}

\section{Training Experiments}
\label{sec:training}

% Our first set of experiments is designed to evaluate how LibriQuote can be used to fine-tune or train from scratch a baseline TTS system.
% Then, we analyze the performance of state-of-the-art TTS systems on LibriQuote-\textit{test}.
% Then, we evaluate how a baseline TTS system fine-tuned on LibriQuote-\textit{train} quotations behaves on other benchmark datasets.

% \subsection{Training Experiments}

In this section, we analyse the impact of using LibriQuote as a resource for fine-tuning or training TTS systems from scratch.
Our goal is to understand 1) how LibriQuote could be used as an alternative to standard TTS datasets such as Emilia \cite{he2024emilia}, even though it contains 20$\times$ less data and 2) how it can be used to enhance expressivity of pre-trained TTS systems when used as a fine-tuning resource.
% We evaluate how quotations in LibriQuote-\textit{train} (denoted as $\mathbf{Q}$) and the high-expressivity split $\mathbf{Q}_f$ can be used to enhance TTS systems by fine-tuning SparkTTS \cite{sparktts} on these corpora.
We mainly conduct experiments with SparkTTS and F5-TTS, that we chose as baselines for better reproducibility.
SparkTTS is an autoregressive TTS model based on Qwen2-0.5B, that generates global and semantic discrete audio tokens that can be converted back to raw audio.
% Global tokens capture speaker-specific details and semantic tokens derive utterance-specific semantic information.
It was trained on approximately 100K hours of speech data, containing in-the-wild and audiobook corpora.
F5-TTS is a non-autoregressive flow-matching based TTS system, trained on 100K hours of speech data.
We describe SparkTTS and F5-TTS in details in Appendix~\ref{app:baselines}.
For each models, we use the available checkpoints and do not modify generation parameters.
For SparkTTS, we compute global and semantic tokens for each training utterances,
and fine-tune the LLM-backbone using the standard language modeling task on semantic tokens with text and global tokens prepended in the sequence.
For F5-TTS, we use the provided fine-tuning script available in the official repository\footnote{\url{https://github.com/swivid/f5-tts}}.
Training details including optimizer and hyperparameters can be found in Appendix~\ref{app:sparktts} and Appendix~\ref{app:f5}.
% on the two corpora (SparkTTS$_{\text{FT}, Q}$ and SparkTTS$_{\text{FT}, Q_f}$).
% (details can be found in the Technical Appendix).
% We employ Adam optimizer \cite{kingma2017adammethodstochasticoptimization} with peak learning rate of 1e-5.
% We train each model for 3 epochs on their respective data.
% A cosine schedule is used with 10000 warmup steps for the model fine-tuned on $\mathbf{Q}$ ($\text{FT}(\mathbf{Q})$) and 5000 steps when using $\mathbf{Q}_f$ ($\text{FT}(\mathbf{Q}_f)$).
% Note that $\mathbf{Q}_f$ contains $10\times$ less utterances than $\mathbf{Q}$).
% In each scenario, we form batches of 32 utterances.
% During inference, we compute global tokens using reference narration utterances, and prepend the text tokens and global tokens before starting the generation, following SparkTTS setup.

We design training experiments in two ways: fine-tuning experiments with varying quotation dataset (either the full set of quotations $\mathbf{Q}$ or the expressive subset $\mathbf{Q_f}$), and training from scratch experiments with either the full quotation subset $\mathbf{Q}$ or the full training dataset (both narrations and quotations $\mathbf{N} \cup \mathbf{Q}$).
% Then, we train two variants of SparkTTS from scratch using the quotation subset $\mathbf{Q}$, initializing with Qwen2-0.5B weights.
% We initialize SparkTTS with Qwen2-0.5B weights, and continue training by learning to autoregressively predict the next semantic tokens, conditioned on the text and global tokens.
Besides, we experiment with replacing SparkTTS text-condition with contextual information in the vicinity of a target quotation (denoted as \texttt{Ctxt}). By having access to more contextual information, we expect the second variant to improve expressivity over standard SparkTTS. More details can be found in Appendix~\ref{app:sparktts}.

\begin{table}[t!]
\centering
\small
\setlength{\tabcolsep}{4.5pt}
\begin{tabular}{l|cc|cc}
\toprule
% & \multicolumn{2}{c}{\textbf{LibriSpeech PC}} &  \multicolumn{2}{c}{\textbf{SeedTTS}-\textit{test-en}} \\
 & \multicolumn{2}{c}{\makecell{\textbf{LibriSpeech} \\ \textbf{PC}}} & \multicolumn{2}{c}{\makecell{\textbf{SeedTTS} \\ \textit{\textbf{test-en}}}} \\
 \midrule
 & \textbf{WER} $\downarrow$ & \textbf{SIM} $\uparrow$ & \textbf{WER} $\downarrow$ & \textbf{SIM} $\uparrow$  \\
 \midrule
Ground Truth & 2.44 & 0.69 & 2.06 & 0.73 \\
\midrule
\midrule
% CosyVoice & 3.59 & 0.66 & 3.39 & 0.64 \\
% F5-TTS & 2.42 & \textbf{0.66} & \textbf{1.83} & 0.67 \\
% % MaskGCT & - & - & 2.62 & \textbf{0.72} \\
% \midrule
SparkTTS & 3.06\phantom{$^\star$} & \textbf{0.52}\phantom{$^\star$} & 2.64\phantom{$^\star$} & 0.46\phantom{$^\star$} \\
\midrule
% F5-TTS & 2.42 & 0.66 & 1.83 & 0.67 \\
% \midrule
    % SparkTTS$_{\text{Scratch}}$ & 3.26 & 0.48 & 5.14 & 0.41 \\
    {\text{FT} ($\mathbf{Q}_f$)} & 2.10$^\star$ & 0.51\phantom{$^\star$} & 2.07\phantom{$^\star$} & 0.42$^\star$  \\
    {\text{FT} ($\mathbf{Q}$)} & \textbf{2.00}$^\star$ & 0.51\phantom{$^\star$}  & 1.90$^\star$ & 0.42$^\star$ \\
    Scratch ($\mathbf{Q}$) & 3.27\phantom{$^\star$} & 0.47$^\star$ & 5.14$^\star$ & 0.41$^\star$\\
    Full ($\mathbf{N} \cup \mathbf{Q}$) & 2.22$^\star$ & 0.49$^\star$ & 4.05$^\star$ & 0.41$^\star$\\ 
    \;\; + FT  ($\mathbf{Q}_f$)  & 2.49$^\star$ & 0.49$^\star$ & 4.23$^\star$ & 0.42$^\star$ \\ 
\midrule
\midrule
F5-TTS & 2.11 & 0.66 & 1.69 & 0.67  \\
\midrule 
    {\text{FT} ($\mathbf{Q}_f$)} & 2.13& 0.66 & \textbf{1.64} & \textbf{0.67} \\
\bottomrule
\end{tabular}
\caption{Word Error Rate (WER) and speaker similarity (SIM) on LibriSpeech-PC and Seed-TTS \textit{test-en}. Two-sided paired student $t$-tests are conducted against SparkTTS ($^\star$ indicates $p<0.05$).}
\label{tab:benchmark}
\end{table}

\subsection{Metrics}
\label{sec:metrics}

Our evaluation follows the \textit{cross-sentence} design \cite{le2023voicebox}: a narration utterance of 2 to 15 seconds is used as audio context to guide the generation of a quotation utterance.

\paragraph{Objective Metrics} We report WER computed with Whisper-large-v3 \cite{radford2023robust} to measure speech intelligibility and speaker similarity between the synthesized speech and the original ground-truth speech (SIM-O). We employ a WavLM-large based speaker verification model \cite{chen2022large} to extract speaker embeddings and calculate cosine similarities.
% We evaluate prosody similarity with the ground truth quotation using Mel Cepstral Distortion (MCD) and F0 Pearson correlation (FPC) following \cite{huang2023singing}.
We evaluate emotion similarity (E-Sim) by leveraging representations from Emotion2Vec \cite{ma-etal-2024-emotion2vec} (\texttt{emotion2vec-plus-base}).
% and an ECAPA-TDNN model trained on Common Voice \cite{desplanques20_interspeech, ardila-etal-2020-common, zuluaga2023commonaccent}.
Similarity scores are calculated by computing the cosine similarity between the ground-truth and synthesized speech representations.
% \footnote{\url{https://huggingface.co/emotion2vec/emotion2vec_plus_base}} 
% \footnote{\url{https://huggingface.co/Jzuluaga/accent-id-commonaccent_ecapa}} 

\begin{table*}[t!]
\centering
\small
\setlength{\tabcolsep}{5pt}
\begin{tabular}{l|ccc|cc|cc}
\toprule
 & \textbf{WER} $\downarrow$ & \textbf{SIM-O} $\uparrow$  & \textbf{E-Sim} $\uparrow$ & \textbf{ContextMOS} & \textbf{Win-Rate} & \textbf{CMOS} & \textbf{MOS} \\
 \midrule
Ground Truth & 6.5 & - & $0.62^\star$ & 3.55 \tiny{$\pm 0.20$}  & - & 0.0 & 3.58 \tiny{$\pm 0.15$}\\
\midrule
SparkTTS & \textbf{4.8} & 0.46 & 0.69  & 2.94 \tiny{$\pm 0.20$} & 38\% & -0.97 \tiny{$\pm 0.25$} & 3.26 \tiny{$\pm 0.15$}\\
F5-TTS & 6.9 & 0.53 & 0.71  & 2.95 \tiny{$\pm 0.21$} & 31\% & -0.98 \tiny{$\pm 0.25$} & 3.47 \tiny{$\pm 0.16$}\\
MaskGCT & 7.6 & \textbf{0.56} & \textbf{0.72}  & 2.94 \tiny{$\pm 0.20$} & 28\%  & -0.98 \tiny{$\pm 0.22$} & 3.36 \tiny{$\pm 0.16$}\\
IndexTTS2 & 5.2 & 0.49 & 0.63  & 3.33 \tiny{$\pm 0.19$} & 46\% &  \textbf{0.25} \tiny{$\pm 0.42$} &  3.47 \tiny{$\pm 0.29$} \\
IndexTTS2-\textit{Context} & 5.4 & 0.50 & 0.64  & \textbf{3.45} \tiny{$\pm 0.21$} & \textbf{54\%} & 0.03 \tiny{$\pm 0.45$} &  \textbf{3.63} \tiny{$\pm 0.29$}\\

\bottomrule
\end{tabular}
\caption{Zero-shot TTS results on LibriQuote-\textit{test} set. Similarity metrics are computed against Ground Truth ($^\star$ indicates similarity against reference). Bootstrapped 95\% confidence intervals are reported with $\pm$.}
\label{tab:res1}
\end{table*}

\paragraph{Contextual Metrics} To measure to which extent synthesized speech matches the quotation context, we leverage recent advances in Large-Audio-Language-Models (LALMs) and use an LALM-as a-Judge approach \cite{manku2025emergentttsevalevaluatingttsmodels,ji2025wavrewardspokendialoguemodels}.
We follow \citet{manku2025emergentttsevalevaluatingttsmodels} and use Gemini-2.5 Pro as the Judge model, as it was proved to correlate strongly with human judgements.
We build an expressive test-subset containing 201 quotations from LibriQuote-\textit{test}, sampling only quotations that are matched with a predicted adverb, following the methodology described in Section~\ref{sec:narrinfo}.
We evaluate synthesized and ground-truth utterances in two ways: a single sample Context Mean Opinion Score (ContextMOS) and a two-sample comparative Win-Rate score as defined in \citet{manku2025emergentttsevalevaluatingttsmodels}.
ContextMOS rates on 1-5 scale how an utterance speaking style matches the book context, while Win-Rate calculates the percentage of times a synthesized utterance delivers a more appropriate speaking style than a ground-truth sample, based on the book context.
% Given a quotation text, its book context and an audio sample, ContextMOS prompts Gemini-2.5 Pro to assign a score between 1 and 5, where 1 indicates complete failure at expressing the appropriate speaking style (e.g. \textit{whispering}), and 5 indicates natural delivery.
% To compute Win-Rate scores, we prompt Gemini-2.5 Pro with a ground-truth quotation sample and an associated synthesized sample, and ask to provide a contextual match score from a scale from 1 to 3 for each sample individually, and then chose which sample conveys the most appropriate speaking style based on these individual scores.
% Win-Rate is calculated as the percentage of time a model wins over ground-truth samples.
% In each case, we require an extensive reasoning trace before providing a score.
Details can be found in Appendix~\ref{app:metrics}.

\subsection{Benchmark Datasets}

We leverage standard benchmark datasets to evaluate in-domain and out-of-domain TTS performance.
We use LibriSpeech-PC \cite{librispeechpc}, a punctuation restored version of LibriSpeech with the standard split from \citet{f5tts} that contains 1127 samples with 4-to-10 seconds audio prompts and SeedTTS \textit{test-en} \cite{anastassiou2024seed}, which is composed of 1088 samples from Common Voice \cite{ardila-etal-2020-common}.
% We do not report scores for the Mandarin variant of SeedTTS as our fine-tuning data only support English speech.
% , although it would be interesting to see how fine-tuning impacts performance on other languages.
% These two datasets are standard TTS benchmarks that we use to measure the impact of using LibriQuote as training data on different in-domain (LibriSpeech-PC) and out-of-domain (SeedTTS) data.
We report WER and cosine similarity between synthesized speech and reference sample (SIM) with the same models as described above.

\subsection{Analysis}

Table~\ref{tab:res2} displays results on LibriQuote-\textit{test}.
The ground-truth WER is relatively high compared to standard audiobook datasets such as LibriSpeech. As discussed in Section \ref{sec:desc_stats}, this is partly due to strong emotions or accents in some samples, which increases recognition errors.
In contrast, SparkTTS and its variants fine-tuned with $\mathbf{Q}_f$ and $\mathbf{Q}$ achieve lower WER than the ground truth. These fine-tuned models show slightly higher SIM-O and E-Sim, while ContextMOS and win rates remain similar, suggesting that fine-tuning on LibriQuote modestly improves intelligibility and speaker similarity but not contextual expressiveness.
Conversely, models trained from scratch show the opposite trend: using standard text conditions yields a ContextMOS of 3.09 when using the quotation subset as training data, which increases to 3.15 when book context is used as input.
This improved expressivity comes with reduced intelligibility and speaker similarity, likely due to a limited amount of training data.
This decrease in intelligibility is drastically reduced when using the both narrations and quotations ($\mathbf{N} \cup \mathbf{Q})$ when training from scratch.
This model achieves greater expressivity compared to all SparkTTS baselines with a ContextMOS of 3.30.
Interestingly, fine-tuning this model on a second-stage with the expressive subset $\mathbf{Q_f}$ does not yield significant gains, echoing the other fine-tuning experiments.

Compared to SparkTTS, we found that fine-tuning conducted with F5-TTS is able to yield large expressivity gain.
When fine-tuning with the filtered quotation subset $\mathbf{Q_f}$, F5-TTS achieves better speech intelligibility and improved expressivity, with ContextMOS reaching 0.4 points higher.
This notable difference is likely related to the different training objectives: while SparkTTS uses an autoregressive loss, it seems that the flow-matching loss of F5-TTS is better able to pick the important expressive information present in LibriQuote's quotations.

\paragraph{Benchmark Datasets} In Table~\ref{tab:benchmark}, we present results for benchmark datasets.
We omit the from-scratch model with context because these datasets lack additional contextual information.
SparkTTS results are reported using samples generated by our synthesis pipeline to ensure fair comparison.
Unlike LibriQuote-\textit{test}, fine-tuned models achieve lower WER than SparkTTS on LibriSpeech-PC and SeedTTS-\textit{test-en}, indicating better intelligibility. We attribute this to fine-tuning acting as a data curriculum, where high-quality data is introduced late in training, a strategy used in recent TTS training strategies \cite{atamanenko2025tts1technicalreport}.
On LibriSpeech-PC, SIM is comparable between SparkTTS and FT models, but degrades on the out-of-domain SeedTTS-\textit{test-en}. The Scratch model shows similar WER to SparkTTS on LibriSpeech-PC but performs much worse on SeedTTS, as expected given the mismatch between audiobook and in-the-wild speech.
The SparkTTS model trained on the full dataset ($\mathbf{N} \cup \mathbf{Q})$ largely improves on speech intelligibility and similarity over the other from-scratch variant, but still struggles on out-of-domain data.
Further fine-tuning this model on the expressive subset $\mathbf{Q_f}$ slightly decreases speech intelligibility on both datasets, further suggesting that fine-tuning is not necessary for a model already trained on the full LibriQuote dataset. 

These experiments conclude that supervised fine-tuning on LibriQuote significantly enhances SparkTTS intelligibility on in-domain and out-of-domain data, and that training from scratch enhances expressivity at the cost of less intelligible speech.
Besides, fine-tuning a different TTS backbone yields a different picture, showing improves expressivity and intelligibility for F5-TTS.
This difference highlights important research directions in understanding the reasons behind less performant SparkTTS fine-tuned models.

% This suggest that further research on different TTS architectures and mixtures of LibriQuote quotations and other data (such as neutral narrations that we did not use in our experiments) could reveal potential breakthrough in expressive TTS.
Thus, LibriQuote appears as a promising resource to fine-tune flow-matching based TTS systems to be more expressive, or to train from-scratch autoregressive TTS models for an enhanced expressivity.
% but requires additional research to find a balance between intelligible and expressive speech.

% We omit the model trained from scratch with context since these datasets do not provide additional contextual information.
% We report SparkTTS performance from samples generated with our synthesizing pipeline to ensure comparable results across variants.
% Differently from LibriQuote-\textit{test}, fine-tuned models exhibit lower WER than SparkTTS on both LibriSpeech-PC and SeedTTS-\textit{test-en}, indicating more intelligible speech.
% We hypothesize that this fine-tuning step can be seen as a sort of Data Curriculum where high-quality data is provided at the end of training -- a strategy now present in Speech Language Model fine-tuning stages \cite{atamanenko2025tts1technicalreport} -- improving intelligibility on neutral and in-the-wild speech.
% While SIM is not significantly different between SparkTTS and fine-tuned variants on LibriSpeech-PC, we still observe a drop in speaker similarity on out-of-domain SeedTTS-\textit{test-en}.
% Besides, the model trained from scratch shows a WER that is not significantly different from SparkTTS on LibriSpeech-PC, but that is way larger on SeedTTS.
% This is expected as Audiobook speech differs in terms of content from SeedTTS's in-the-wild speech.

\section{Benchmarking with LibriQuote-\textit{test}}
\label{sec:benchmark}

In this section, we evaluate the ability of state-of-the-art TTS systems to synthesise speech as expressive as amateur audiobook readers from LibriQuote-\textit{test}.
We focus our evaluation on aspects of speech naturalness and emerging capacities to systematically deliver speech utterances that satisfy an implicit or explicit speech intent (e.g. ``whispering'' or ``happiness'').
We compare four recent open-source TTS systems: SparkTTS \cite{sparktts}, F5-TTS \cite{f5tts}, MaskGCT \cite{maskgct} and IndexTTS2 \cite{zhou2025indextts2breakthroughemotionallyexpressive}.
% To evaluate to which extent state-of-the-art TTS systems are able to synthetize expressive speech, we compare four recent open-source TTS systems: SparkTTS \cite{sparktts}, F5-TTS \cite{f5tts}, MaskGCT \cite{maskgct} and IndexTTS2 \cite{zhou2025indextts2breakthroughemotionallyexpressive}.
These models were chosen based on their availability and performance on standard benchmarks, and are described in Appendix~\ref{app:baselines}.
% SparkTTS is an autoregressive TTS model based on Qwen2-0.5B.
% It generates global and semantic discrete tokens can be converted back to raw audio.
% Global tokens capture speaker information such as timbre, while semantic tokens capture fine-grained semantic information.
% F5-TTS is a non-autoregressive flow matching TTS system based on Diffusion Transformer \cite{dit}.
% MaskGCT is a non-autoregressive TTS system, that follows a \textit{mask-and-predict} paradigm.
% IndexTTS2 is an autoregressive flow matching TTS system that enables fine-grained control over style and emotions.
Notably, IndexTTS2 first predicts a 7-class emotion distribution based on an emotion prompt (with the text to synthesize as default), that conditions the synthesized speech.
This decoupled approach is well-suited for LibriQuote, as the emotion prediction can be done on contextual information around quotations rather than quotation-text alone.
We denote this approach as IndexTTS2-\textit{Context}.
% It follows a two-stage process: semantic tokens derived from a VQ-VAE model with Wav2Vec-Bert 2.0 features are predicted in a first step, and then used to condition the prediction of acoustic tokens.
All models were trained with the Emilia dataset \cite{he2024emilia} on approximately 50K to 100K hours of speech data, but SparkTTS and IndexTTS2 also use additional Audiobook speech data.
We use publicly available checkpoints for each model and do not modify generation parameters.
% Baselines details are available in Appendix~\ref{app:baselines}.

\subsection{Metrics}

We use the same set of objective and contextual metrics defined in Section~\ref{sec:metrics}.
In addition, we conduct a subjective rating with human raters.
We employ Mean Opinion Score (MOS, rated from 1 to 5 with 0.5 intervals) to assess naturalness, and Comparative MOS (CMOS, rated from -3 to 3) to measure the degree of expressivity with respect to the ground-truth.
We randomly selected 2 samples per speaker in LibriQuote-\textit{test}, for a total of 30 samples.
Each sample was judged by 5 raters.
More details can be found in Appendix~\ref{app:subj_metrics}.

\subsection{Analysis}
% As shown in Table~\ref{tab:res1}, the ground-truth WER is relatively high compared to standard audiobooks datasets such as LibriSpeech.
% As discussed in Section~\ref{sec:desc_stats}, some samples exhibit large WER because of strong emotion or accents that leads to higher recognition rate errors.
Results are displayed in Table~\ref{tab:res1}.
SparkTTS has the lowest WER (4.8) but the lowest SIM-O, while MaskGCT shows the highest WER and SIM-O.
F5-TTS and MackGCT produce speech with relatively high emotion similarity.
Both IndexTTS2 variants achieve low WER and high ContextMOS and Win-Rates, with IndexTTS2-\textit{Context} matching ground truth in Win-Rate (54\%).
Subjective tests further show IndexTTS2 reaches positive CMOS, meaning it is rated more expressive than ground truth on average.
Other TTS systems sound natural (high MOS) but fail to capture appropriate prosody, as indicated by negative CMOS and low ContextMOS, highlighting IndexTTS2’s clear advantage in expressive speech synthesis.

% SparkTTS achieves the lowest WER (4.8) across models, but the lowest SIM-O.
% In contrast, MaskGCT achieves the highest WER and SIM-O.
% F5-TTS and MackGCT synthesize speech with higher emotion similarity than other models.
% Interestingly, both variants of IndexTTS2 achieve relatively low WER, and high ContextMOS and Win-Rates.
% IndexTTS2-\textit{Context} even matches ground-truth samples in Win-Rate (53\%), indicating that it is able to synthesize utterances with contextual speaking style as good as humans on our test-subset.
% This result is also supported by subjective experiments, where IndexTTS2 reaches a positive CMOS, indicating that it was rated more expressive than ground-truth quotations by human raters on average.
% In contrast, other TTS systems seem to generate natural speech (high MOS), but  fail at synthesizing the appropriate prosody as shown by a consistent negative CMOS and low ContextMOS, suggesting a clear gap between IndexTTS2 and other baselines in terms of synthesizing highly expressive speech.

While IndexTTS2 already captures emotional expressivity well, replacing text conditioning with contextual conditioning for emotion prediction further improves contextual metrics, achieving human-level Win-Rate ($\geq$50\%) and MOS naturalness.
These results show that decoupling emotion prediction from speech synthesis significantly enhances contextual expressivity.
Although IndexTTS2 generates more expressive speech than other baselines, it also yields the lowest E-Sim, likely due to emotion prediction errors from text alone. Incorrect emotion predictions can produce speech that contradicts the intended style, which might mislead listeners in applications such as Audiobook reading. 
As detailed in Appendix~\ref{app:context_match}, Quotations read by amateur speakers do not always convey sufficient emotion, but still render the appropriate style more often than all models.
In contrast, IndexTTS2 exhibits a higher proportion of both appropriate and mismatched speaking styles than other models, an issue partially mitigated by IndexTTS2-Context.

% While classic IndexTTS2 captures quite well emotional expressivity, we found that replacing the text-condition by the quotation's context to predict the emotion distribution (IndexTTS2-\textit{Context}) improves performance on contextual metrics, reaching human parity on Win-Rate ($\geq 50\%$) and MOS naturalness.
% These results show that decoupling emotion-prediction from speech synthesis, as done in IndexTTS2, largely improves sample contextual expressivity.
% Interestingly, while IndexTTS2 seamlessly produce more expressive speech than other baselines, it also achieves the lowest E-Sim across baselines, which we think might be due to potential emotion prediction errors.
% Predicting the wrong emotion distribution from text can lead to a potential higher proportion of speech that is in complete contrast with the emotion expected from the context.
% Indeed, we analysis this aspect in Appendix~\ref{app:context_match} and show that IndexTTS2 provides a larger proportion of both appropriate and wrong speaking styles than other models, that is, however, largely improved by IndexTTS2-\textit{Context}.

\section{Conclusion}

We proposed LibriQuote, a public speech dataset derived from audiobooks, that includes 12,720 hours of neutral narration and 5,359 hours of expressive quotations from fictional characters read by amateur readers.
Our qualitative analysis reveals that LibriQuote-\textit{test} is more emotionally diverse than previous audiobook corpora.
Besides, each utterance of LibriQuote is supplemented with its book context and pseudo-labels of quotation intent, showing a wide-range of intended speaking style.
Fine-tuning experiments show both training subsets improve SparkTTS intelligibility on other corpora while training from scratch on the full quotation set enhances expressiveness, and that fine-tuning F5-TTS on LibriQuote's expressive subset yields substantial expressivity gains.
Evaluation on LibriQuote-\textit{test} shows that IndexTTS2 matches ground-truth expressivity, though some synthesized quotations convey unintended emotions.
In-depth analysis revealed that amateur readers sometimes convey insufficient emotion in their rendering of quotations, but TTS systems do so at an even higher proportion.
% in terms of rendering of expressive speech.
Despite recent advancements, TTS systems require substantial improvements to meet standards of professional audiobook narration.

We also want to highlight LibriQuote's value for practitioners of Digital Humanities.
LibriQuote provides 3 million character quotations in text and speech modalities, from 2,991 public domain novels, with pseudo-labels of quotation intent.
This large-scale data offers new perspectives to analyse Audiobook prosody, e.g. comparing how male/female speakers interpret male/female characters \cite{Pethe_2025} or understanding stylistic differences of words uttered by characters \cite{michel-etal-2024-distinguishing}. 
% This large-scale dataset enables analysis of audiobook prosody, including gendered character portrayal, alignment of speech with context, and stylistic differences in character dialogue.

% While we validated LibriQuote through an extensive TTS evaluation, we also note its potential analytical value for practitioners of Digital Humanities.
% LibriQuote offers more than 3 millions character quotations (in both text and speech modalities) across 2991 public domain fiction novels drawn from Project Gutenberg, supplemented with pseudo-labels of direct speech cues describing the quotation intents.
% This unprecedented scale of speech data delivers new perspective to analyse Audiobook prosody, e.g. comparing how male/female speakers portray male/female characters \cite{Pethe_2025}, identifying alignment between quotation context and delivered speech, or understanding stylistic differences in words uttered by characters \cite{michel-etal-2024-distinguishing, vishnubhotla-etal-2019-fictional}. 
% TTS systems we evaluated fail to synthesize speech as expressive as the ground-truth utterances and that LibriQuote-\textit{train} improves speech intelligibility when used as a fine-tuning corpus.
% We believe that the release of this dataset can foster research on expressive TTS and on contextual conditioning for audiobook synthesis.

\section{Limitations}

This work bears several limitations.
First, we did not filter LibriQuote-\textit{train} utterances that exhibit very high WER between the transcript and ground-truth text as done in previous work.
Thus, LibriQuote might contain a small number of outliers, which can hinder learning of ASR or TTS systems.
We plan to perform WER-based filtering in a future dataset release.

% Second, the narrative information extracted with Phi-4 is not perfect.
% Although we maximized precision in order to avoid hallucinations, it might be that a small amount of quotations are assigned an expressive speech verb or adverb wrongly.
% Second, our objective analysis of expressiveness remains limited, with metrics such as FPC and MCD that assume a ground-truth sample is deterministic, hence penalizing diversity in generation.

Second, our evaluation of expressivity relies on an LALM-as-a-judge approach, which is inherently biased by the LALM understanding abilities and own cultural biases.
While prior works have incorporated such approaches in their evaluation pipelines, we believe these automated results should mostly be used to compare models against each other rather than make individual conclusions about models.
Our findings, however, reveal closely aligned trends between human and model judgments, indicating that IndexTTS2 is a substantially more capable TTS system for producing expressive speech.

% Third, training experiments were solely conducted with SparkTTS, which bears its own limitations (such as low speaker similarity), and did not involve other system types.
% Besides, we completely omitted the 12K hours of neutral narration in all training experiments, focusing solely on quotations.
% However, we believe that these baseline system results are still a valuable source of information. 

In addition, although our experiments focused directly on end-to-end TTS, we still note that LibriQuote ; and in particular its filtered subset ; might be well-suited for the training of Neural Audio Codecs.
Indeed, numerous Audio Codecs targeting English language use LibriSpeech or LibriTTS as training data \cite{sparktts, zhang2024speechtokenizerunifiedspeechtokenizer}.
We thus believe that exploring the use of LibriQuote as training data for Neural Audio Codecs offers an interesting area of research for higher-fidelity reconstruction of expressive speech.

% , that might lead to future research.
Finally, we experimented with narrative context conditioning in utterance synthesis, but we believe further research should be conducted on how TTS systems may benefit from it.

\section{Ethical Considerations}

This work is purely dedicated to research projects.
% We acknowledge that releasing datasets designed to improve expressiveness of TTS systems might have considerable negative societal impact.
While TTS models continue bridging the gap towards human naturalness and expressiveness, numerous dangerous applications arise, such as voice spoofing, or the forced replacement of talented voice actors by AI narrators.
Therefore, we believe it is of paramount importance to ensure responsible and ethical applications of TTS.

We also note that LibriQuote does not provide explicit gender or accent distributions of its speakers.
As a results, TTS models trained with LibriQuote are likely to exhibit gender or accent bias ; where one gender or one/multiple accents are either generated more faithfully or generated with stereotypical cues.
We think researchers and/or companies using LibriQuote as training data in the purpose of creating a commercial tool should systematically assess and, if possible, mitigate such bias before providing users with an end product.

The human study involved participants recruited on a voluntarily basis from our organization, and we ensure each listening test was completed within working hours.

\section{Acknowledgements}
This work was performed using HPC resources from GENCI–IDRIS (Grant AD011011668R5)

\bibliography{bib}

% \bibliographystyle{acl_natbib}
% \newpage

\appendix

\section{LibriQuote Analysis}
\label{app:emo}

\begin{figure*}
    \centering
    \includegraphics[width=0.9\linewidth]{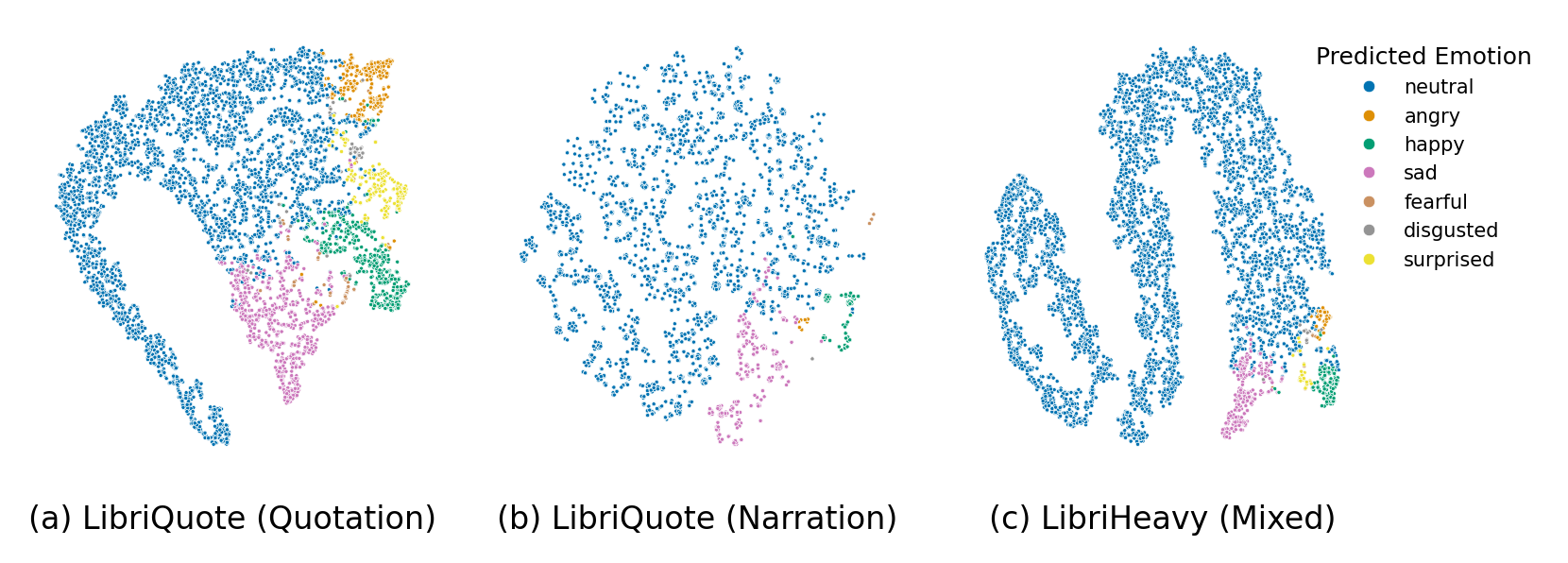}
    \caption{t-SNE projections of utterance representations computed with \texttt{emotion2vec-plus-base}.}
    \label{fig:tsne}
\end{figure*}

As mentioned in Section~\ref{sec:desc_stats}, we performed emotion detection with the open source Emotion2Vec-plus-base.
We report the detection results in Table~\ref{tab:emo_prop} and display t-SNE projections of the resulting utterance representations in Figure~\ref{fig:tsne}, and compare LibriQuote-\textit{test} quotations, narrations and LibriHeavy segments.
As expected, we see a much higher proportion of quotations predicted as non-neutral in LibriQuote quotations compared to LibriHeavy and narration segments.
% We also display in Figure~\ref{fig:accents} t-SNE projections of accent representations computed with a dedicated model.

\section{Narrative Information}
\label{app:narr_info}

\subsection{Prompting Experiments}
% \label{app:6}

We display in Figure~\ref{fig:prompt} the prompt used to extract narrative information for each quotation.
Note that we masked in-context quotations with special markers, and replace the target quotation with ``[TARGET]''.
This is to ensure that only narrative elements remain in the contextual surroundings of a target quotation.

Additionally, we tested various Large Language Models (LLMs) for the task of narrative information extraction, and also tested the extraction of other elements such as \textit{nouns} and \textit{adjectives}.
Full results can be found in Table~\ref{tab:narr_extract}.
Nouns and adjectives often occur when the utterance description is supplemented by additional information (\textit{e.g.} ``he said with a \textit{sigh}'' for nouns and ``he added in a \textit{loud}, \textit{sharp} tone'' for adjectives). 
Note that to ensure a high-quality filtering, we tried to maximize the precision of predictions in order to avoid potential hallucinations where some narrative information are spuriously associated to utterances.
% We see in Table~\ref{tab:prompt_more} that all tested models seem to perform relatively well on verbs, but some across-model variance is manifested for adverbs.
We found that Qwen2.5-72b with self-confidence (SC) provides the best overall performance for these two types of information, followed closely by Phi-4 with SC.
In the end, we decided to use Phi-4 with SC to extract information on the full train set, as it obtained very high precision in 10$\times$ less computation time than Qwen2.5-72b with SC.
Besides, we found that LLMs dramatically fail at extracting accurately adjectives and nouns, with very low F1 scores.
We found during the annotation that extracting such information is harder, even for humans, as introducing adjectives and nouns is less grounded in linguistic rules, and authors might use a more diverse set of linguistic devices to introduce these information.
Thus, we decided not to rely on the extraction of such elements.

Overall, extracting verbs and adverbs with Phi-4 with SC (we used vLLM \cite{kwon2023efficient} as the inference backend) on the full train set took around 6 hours on a single NVIDIA H100 GPU card.

\begin{table}[t!]
    \centering
    \small
    \setlength{\tabcolsep}{4pt}
    \begin{tabular}{l|ccc}
    \toprule
    & Quotations & Narrations & LibriHeavy \\
    \midrule
    \textit{Neutral} & $66.9\%$ & $86.8\%$ & $91.4\%$\\
    \textit{Angry} & $4.6\%$ & $0.5\%$ & $0.8\%$ \\
    \textit{Sad} & $13.6\%$ & $9.3\%$ & $4.6\%$ \\
    \textit{Happy} & $8.4\%$ & $2.7\%$ & $2.3\%$ \\
    \textit{Surprised} & $4.5\%$ & $0.3\%$ & $0.6\%$ \\
    \textit{Fearful} & $0.9\%$ & $0\%$ & $0.1\%$ \\
    \textit{Disgusted} & $1.1\%$ & $0.4\%$ & $0.3\%$ \\
    \bottomrule
    \end{tabular}
    \caption{Proportion of emotion labels predicted by Emotion2Vec. We excluded \textit{Unknown} and \textit{Other} categories.}
    \label{tab:emo_prop}
\end{table}

\subsection{Filtering}
\label{app:filtering}

% We leveraged the pseudo-labels of verbs and adverbs extracted with Phi-4$_\text{conf}$ to produce the filtered subset $\mathbf{Q}_f$.
The result of the extraction is a (potentially empty) set containing extracted verbs and adverbs from Phi-4$_\text{conf}$.
To produce the filtered subset  $\mathbf{Q}_f$, we leverage independently the extracted verbs and adverbs in the following order: 
\begin{itemize}
    \item[1.] We include all quotations that have a non-empty adverb.
    \item[2.] Using the remaining quotations, we include all utterance that have an extracted verb that fall into a predefined list of speech verbs, $\mathbf{S}$.
    \item[3.] We discard every other quotations.
\end{itemize}
To build the list of speech verbs, $\mathbf{S}$, we started by extracting a large list of 201 potential speech verbs from the web\footnote{https://archiewahwah.wordpress.com/speech-verbs-list/} along with their descriptions.
Then, based on verb descriptions, we discarded every verb that might indicate a \textit{neutral} way of speaking.
The resulting verb list contains 89 speech verbs and can be found in Table~\ref{tab:speech_verbs}.

\begin{table*}[t!]
    \centering
    \scriptsize
    \begin{tabular}{l|c||ccc|ccc|ccc|ccc}
        \toprule
         & Utterance & \multicolumn{3}{c}{Verbs}  & \multicolumn{3}{c}{Adverbs} & \multicolumn{3}{c}{Adjectives} & \multicolumn{3}{c}{Nouns}\\
         \midrule
         Support & 400 & \multicolumn{3}{c}{344}  & \multicolumn{3}{c}{56} & \multicolumn{3}{c}{28} & \multicolumn{3}{c}{14}\\
         \midrule
         &  Runtime (s) & \textbf{P} & \textbf{R} & \textbf{F1} & \textbf{P} & \textbf{R} & \textbf{F1}& \textbf{P} & \textbf{R} & \textbf{F1}& \textbf{P} & \textbf{R} & \textbf{F1}\\
        \midrule
        Llama3.3-70b & 64 & 0.82 & 0.97 & 0.89 & 0.65 & 0.86 & 0.74 & 0.64 & 0.43 & 0.51 & 0.80 & 0.29 & 0.42 \\
        Llama3.3-70b (SC) & - & 0.91 & 0.91 & \textbf{0.91} & 0.8 & 0.54 & 0.65 & 1.0 & 0.05 & 0.09 & 1.0 & 0.07 & 0.13 \\
        % \multicolumn{4}{|c|}{test}
        \midrule
        Phi4 & 6 & 0.85 & 0.93 & 0.89 & 0.74 & 0.81 & 0.77 & 0.59 & 0.48 & 0.53 & 0.45 & 0.36 & 0.32 \\
        Phi4 (SC) & - & \underline{0.92} & 0.86 & 0.89 & \textbf{0.95} & 0.61 & 0.74 & 0.0 & 0.0 & 0.0 & 0.0 & 0.0 & 0.0\\
        \midrule
        Qwen3-32b & 19 & 0.86 & 0.93 & 0.89 & 0.69 & 0.92 & 0.79 & 0.5 & 0.76 & 0.6 & 0.43 & 0.71 & 0.54\\
        Qwen3-32b (SC) & - & \underline{0.92} & 0.89 & \underline{0.9} & 0.81 & 0.85 & \textbf{0.83} & 1.0 & 0.05 & 0.09 & 1.0 & 0.14 & 0.25 \\
        \midrule
        Qwen2.5-72b & 62 & 0.88 & 0.88 & 0.88 & 0.62 & 0.85 & 0.71 & 0.44 & 0.71 & 0.55 & 0.21 & 0.5 & 0.21\\
        Qwen2.5-72b (SC) & - & \textbf{0.98} & 0.8 & 0.88 & \underline{0.93} & 0.73 & \underline{0.82} & 0.67 & 0.19 & 0.25 & 0.38 & 0.21 & 0.16 \\
        \bottomrule
    \end{tabular}
    \caption{Narrative information extraction results for all models and information type. Models with \# parameters $\ge$ 70b are 4-bit quantized. Runtime is calculated in seconds on a single H100 card, with vLLM as the inference backend. Best results for precision and F1 are indicated in bold and second best results are underlined.}
    \label{tab:narr_extract}
\end{table*}

\section{SparkTTS Experiments}
\label{app:sparktts}

Our experiments described in Section~\ref{sec:training} involved fine-tuning and training from scratch numerous variants of SparkTTS with LibriQuote.
We provide details below for complete reproducibility of our experiments.
Whether for fine-tuning or training from scratch, we transform each quotation utterance in the sequence $[T, G, S]$ where $T=(t_1,\dots,t_N)$ are the text tokens, $G=(g_1,\dots,g_{32})$ are SparkTTS global tokens (with a fixed number of 32 tokens) and $S=(s_1,\dots,S_D)$ are semantic tokens, with varying length depending on the utterance length.
Each training experiments involve optimizing the following language modeling task:

$$
L = \frac{1}{D}\sum_{i=1}^D p(s_i|t_1,\dots,t_N,g_1,\dots,g_{32},s_{j<i})
$$
in other words, we optimize the Language Model backbone solely on semantic tokens.

\subsection{Fine-Tuning}

For both fine-tuning experiments, we employ Adam optimizer \cite{kingma2017adammethodstochasticoptimization} with peak learning rate of 1e-5.
We train each model for 3 epochs on their respective data split ($\mathbf{Q}$ and $\mathbf{Q}_f)$.
A cosine schedule is used with 10000 warmup steps for the model fine-tuned on $\mathbf{Q}$ ($\text{FT}(\mathbf{Q})$) and 5000 steps when using $\mathbf{Q}_f$ ($\text{FT}(\mathbf{Q}_f)$).
Note that $\mathbf{Q}_f$ contains $10\times$ less utterances than $\mathbf{Q}$.
In each scenario, we form batches of 32 utterances.
During inference, we compute global tokens using reference narration utterances, and prepend the text tokens and global tokens before starting the generation, following SparkTTS setup, yielding the following input sequence $[T, G_r]$ where $G_r = (g_{1,r}, \dots, g_{32, r})$ are global tokens computed from the reference narration.

\subsection{Training from Scratch}

Here, our experiments include two variants.
The first variant is similar to SparkTTS and uses the text to synthesize as the text-condition, with the unchanged sequence $[T, G, S]$ and training objective $L$.
The second variant leverages contextual information around quotations by replacing the text condition with the sequence $T' =[c^l, start, T, end, c^r]$ where $c^l = (c^l_1, \dots, c^l_n)$ and $c^r = (c^r_1, \dots, c^r_m)$ are the text tokens from the left and right context respectively, $T = (t_1, \dots, t_N)$ are the tokens from the text to synthesize, and $start$ and $end$ are special delimiter tokens.
Note that we also replace each quotation appearing in $c^r$ and $c^l$ with a special token.
Therefore, most of the contextual information can be seen as narrative details.
We use a context of one paragraph for both $c^r$ and $c^l$.
We train each variant for 10 epochs with Adam optimizer, a peak learning rate of 3e-5 and a batch size of 32.
A cosine schedule is used with 20K warmup steps.

\begin{figure*}[t!]
    \centering
    \includegraphics[width=.9\linewidth]{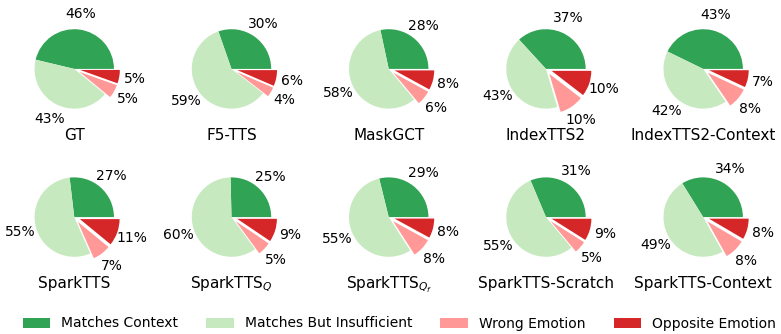}
    \caption{Proportion of failure cases per model predicted by Gemini-2.5 Flash.}
    \label{fig:errors}
\end{figure*}

\section{F5-TTS Experiments}
\label{app:f5}
Our experiments also involve fine-tuning F5-TTS on LibriQuote's expressive subset $\mathbf{Q_f}$.
We use the official fine-tuning script from the f5-tts repository\footnote{\url{https://github.com/SWivid/F5-TTS/blob/main/src/f5_tts/train/train.py}}, starting with the official \texttt{V1-Base} checkpoint.
We use a batch size of 38400 frames per GPU (with 2 H100 GPUs) and a maximum of 32 sequences per batch.
The learning rate is set to 7.5e-5, and we use 20000 warmup steps, training for a total of 3 epochs.

\section{Contextual Metrics}
\label{app:metrics}

With recent breakthroughs in Large Language Models (LLMs) and Large Audio Language Models (LALMs) understanding capabilities, automated evaluations of medium-sized dedicated test samples with LLM-as-a-judge \cite{gu2025surveyllmasajudge} or LALM-as-a-judge \cite{manku2025emergentttsevalevaluatingttsmodels} have become increasingly popular.
These approaches delegate the usual human inspection of some test samples to LLMs or LALMs to provide an aggregated response, which was often proved to correlate strongly with human judges \cite{gu2025surveyllmasajudge, manku2025emergentttsevalevaluatingttsmodels}.
While human judgements would be ideal to evaluate to which extent a quotation sample convey the appropriate emotion and/or voice intent given a particular book context, subjective experiments at scale are expensive in terms of time and costs, and require dedicated knowledge to construct and analyse.
LALM-as-a-judge is thus used as a replacement, noting that a model individual conclusions are often themselves biased, and that only aggregated scores are usually worth exploring.

We follow \citet{manku2025emergentttsevalevaluatingttsmodels} and use Gemini-2.5 Pro as the model judge.
We hypothesize that the combined ability of LALMs to perform audio and text reasoning is valuable in our particular setup, where text reasoning is suited for understanding the right intent to convey for a particular quotation, and audio reasoning for understanding if a sample appropriately exhibits this intent.
We propose two metrics that are built by modifying prompts provided in \citet{manku2025emergentttsevalevaluatingttsmodels}: ContextMOS and Win-Rate.
ContextMOS takes a single audio sample, a quotation text and its context, and prompts Gemini-2.5 Pro to assign a score between 1 and 5, where 1 indicates no alignment between the sample intent and the intents stemming from the context, while 5 indicates complete alignment and naturalness in conveying this intent.
In contrast, Win-Rate takes a ground-truth audio sample and a synthesized sample, a quotation text and its context, and chose which audio sample conveys the right intent the most faithfully.
Note that we allow ties, and define the Win-Rate to of a system $T_i$ relative to the ground-truth as the following:
{
% \small
$$
W(T_i) = \frac{\sum (\text{\scriptsize winner} = \text{\scriptsize  index}_i) + 0.5 \cdot \sum (\text{\scriptsize  winner} = 0) }{n}
$$}

where $0$ indicates ties and $n$ is the number of ratings.
An example prompt for ContextMOS  is provided in Figure~\ref{fig:ctxtmos} and in Figure~\ref{fig:win_rate} for Win-Rate. A similar structure is created for the Win-Rate prompt.
All prompts can be found in the dataset repository.

\section{Baselines}
\label{app:baselines}

In this section, we describe in more details the baselines used in our experiments.

\paragraph{SparkTTS} is an autoregrssive TTS system that builds upon Qwen2.5-0.5M.
It also uses BiCodec, a speech codec model that transforms 16kHz sampling rate speech into two types of discrete tokens: global and semantic tokens.
Global tokens are derived from representations of an ECAPA-TDNN model fine-tuned for speaker-verification, hence capturing speaker-related information such as timbre.
The ECAPA-TDNN representations are mapped to a set of 32 hidden features using learned laten queries, which are then discretized using Finite Scalar Quantization \cite{mentzer2023finitescalarquantizationvqvae}.
Semantic tokens are derived from representations of the 11th, 14th and 16th layers of Wav2Vec 2.0 (XLSR-53) \cite{pasad2023comparativelayerwiseanalysisselfsupervised}.
These features are then quantized using single-codebook vector quantization, using factorized codes as done in DAC \cite{kumar2023highfidelityaudiocompressionimproved}.
Tokens are converted back to raw audio using a Convolutional Neural Network (CNN) Decoder composed of ConvNeXt blocks.
The language modeling framework is used to learn to generate semantic tokens, conditioned on the text-to-synthesis tokens and global tokens (which are prepended in the autoregressive sequence).

\paragraph{F5-TTS} is a non-autoregressive diffusion TTS system based on Flow Matching that does not rely on phonetic alignment nor duration predictor, operating on 24kHz sampling rate speech.
During training, following the text-guided speech infilling task, speech mel spectrograms are first masked and then concatenated with text tokens.
The resulting sequence is inputted to a Diffusion Transformer (DiT) \cite{peebles2023scalablediffusionmodelstransformers}, which learns to generate the masked spectrogram region from gaussian noise using Flow Matching.
During inference, the reference speech mel spectogram serves as the unmasked region, and is concatenated with the text to synthesize the input sequence.
Sway sampling is used along with an Ordinary Differential Equation (ODE) solver to generate mel spectrograms that are converted back using Vocos \cite{siuzdak2024vocosclosinggaptimedomain}.

\paragraph{MaskGCT} is also a non-autoregressive TTS system, that follows a two-step masked generative modeling paradigm.
It leverages two speech codec models to produce semantic and acoustinc discrete units from 24kHz sampling rate speech.
Semantic tokens are derived from representations of the 17th layer of W2v-BERT 2.0, that are converted to discrete units using a single codebook of size 8,192 via classic Vector Quantization (VQ).
Acoustic tokens are produced following RVQGAN \cite{kumar2023high}, and 8 codebooks of size 8,192 are used via Residual VQ (RVQ).
To generate raw audio, MaskGCT follows a two-step process: first, it uses a Transformer to generate semantic tokens by iteratively predicting masked tokens conditioned on the text to synthesize, prompt tokens and previously generated semantic tokens ; second acoustic tokens are generated following a similar iterative procedure, but using an architecture similar to SoundStorm \cite{borsos2023soundstorm}, and the generative process is conditioned with the prompt acoustic tokens and previously predicted semantic tokens.
Finally, predicted acoustic tokens are converted back to raw audio using Vocos.
Additionally, MaskGCT employs a Flow-Matching duration predictor to predict the target speech duration during inference.

\paragraph{IndexTTS2} is a recent autoregressive TTS system that enables fine-grained control over duration and emotions.
It incorporates three modules: a Text-to-Semantic (T2S) module, a Semantic-to-Mel (S2M) module and a vocoder.
The T2S module autoregressively generates semantic tokens derived from a dedicated Neural Speech Tokenizer, trained with the standard language modeling task, while the S2M module is trained to generate faithful mel-spectrogram from semantic tokens, trained with flow matching.
The vocoder converts mel-spectrogram back into raw audio.
The core innovation of IndexTTS2 relies on its capacity to control the duration and emotion of synthesized speech.
In particular, a soft-instruction module derived from Qwen3 predicts an emotion distribution from the text description, that is used to condition the speech synthesis.
This decoupled approach allows innovative disentanglement between emotional expression and speaker identity -- an approach particularly suited for the task of synthesizing character quotation -- leading to large improvements in expressive TTS.

\section{Subjective Experiments}
\label{app:subj_metrics}

Human subjects for the listening tests were recruited from our organization via voluntary participation.
Participants were indicated that they were part of a study involving the comparison of AI-generated speech and human speech.
They were also advised to use headphones and to complete the study in a quiet environment.

For the CMOS experiments, we ask the participants the following question: \textit{How expressive is this recording? Please focus on aspects of style (timbre, emotion and prosody), and ignore the aspects of content, grammar, or audio quality}.
For each instance, we give participants two audios containing one human sample and one generated sample, in a randomized order.
We ask the participants to rate if the first audio sounds more expressive than the second on a $[-3, 3]$ scale.
% Screenshot of the CMOS platform can found in Figure~\ref{fig:cmos_platform}

For the MOS experiments, we ask the participants the following question:
\textit{How natural (i.e human-sounding) is this recording? Please focus on examining the audio quality and naturalness, and ignore aspects of style (timbre, emotion, prosody)}.
For each instance, we give participants a single audio, and ask them to rate the naturalness of the sample on an 1-5 Likert scale with 0.5 intervals.
% Screenshot of the MOS platform can be found in Figure~\ref{fig:mos_platform}

\begin{figure*}
\centering
    \includegraphics[width=.9\linewidth]{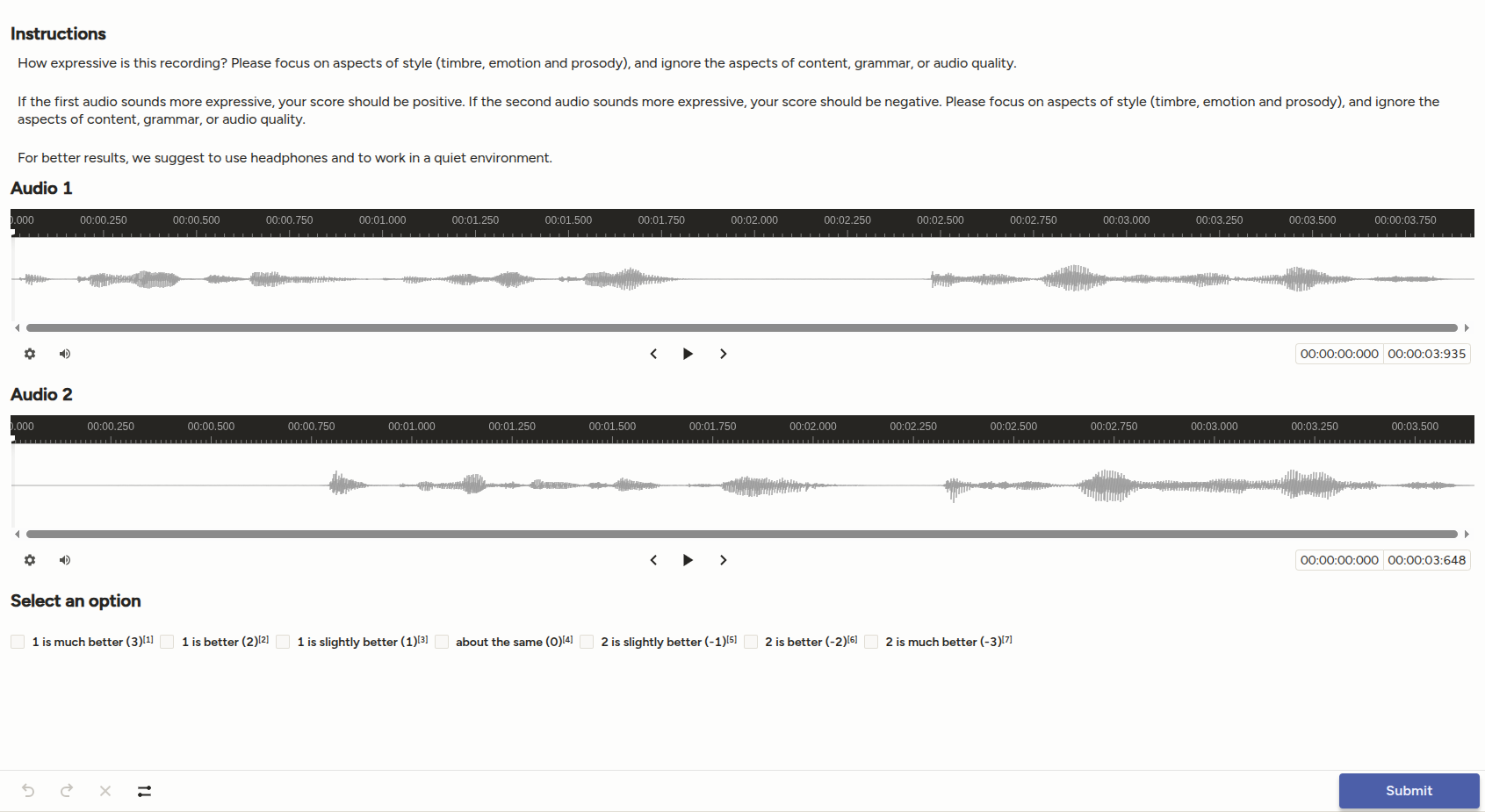}
    \caption{Screenshot of the platform used for the CMOS experiment.}
    \label{fig:cmos_platform}
\end{figure*}

\begin{figure*}
\centering
    \includegraphics[width=.9\linewidth]{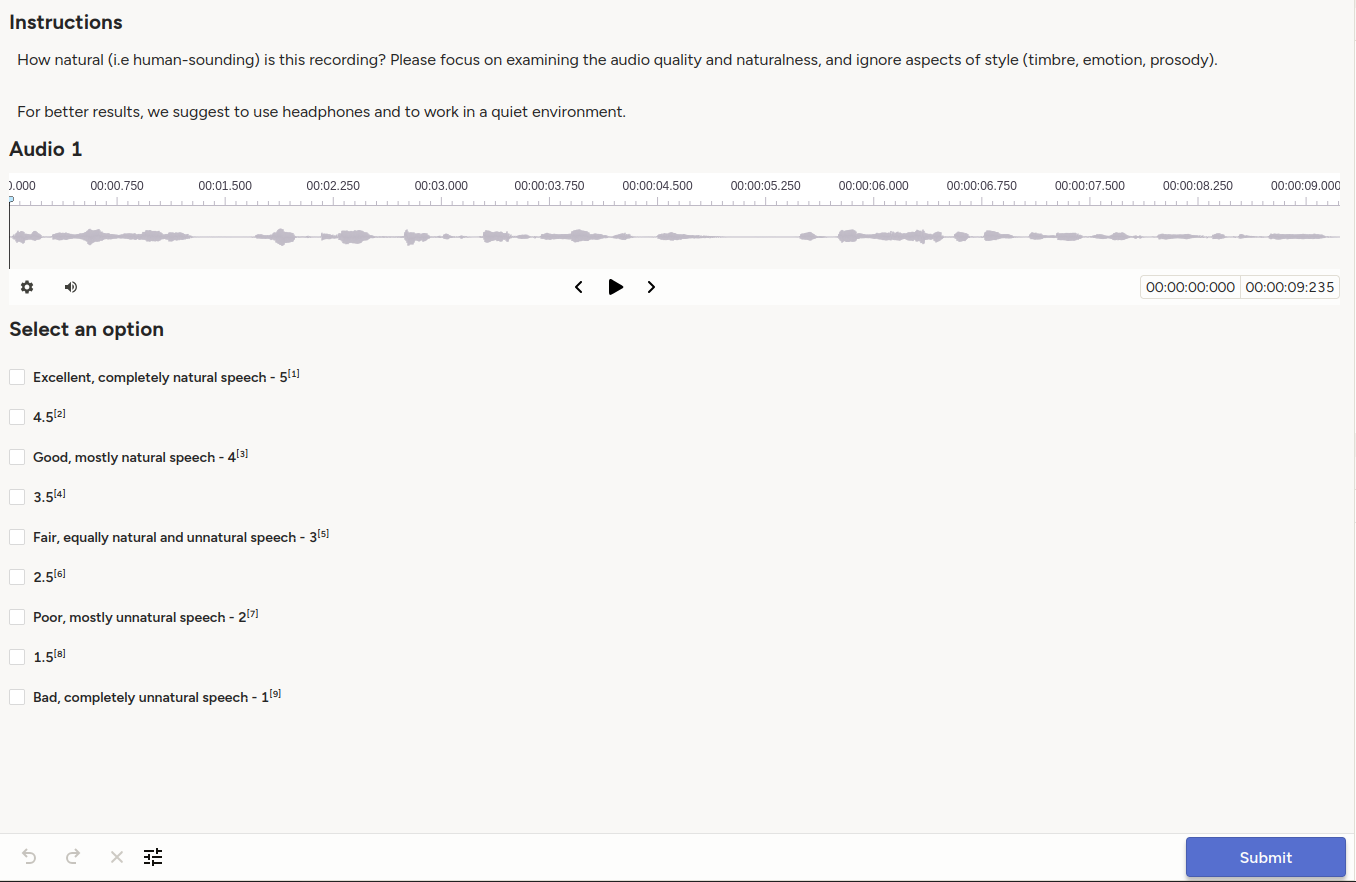}
    \caption{Screenshot of the platform used for the MOS experiment.}
    \label{fig:mos_platform}
\end{figure*}

We provide screenshots of the annotation platform in Figure~\ref{fig:cmos_platform} and Figure~\ref{fig:mos_platform}.

\section{Investigating Contextual Matching}
\label{app:context_match}

We showed that IndexTTS2 is able to produce high-quality expressive speech.
% likely due to its emotion prediction module.
However, we found during listening that some samples could exhibit a very strong emotion that is completely different from what one could expect from the context.
To investigate the proportion of such cases, we leverage the reasoning traces produced by Gemini-2.5 Pro when providing a score for ContextMOS.
These traces contain fine-grained details describing the expected emotion from the context, and what is conveyed appropriately or wrongly in the provided samples, sometimes with precise timestamps.
To provide an overview of failure cases, we provide Gemini-2.5 Flash with a reasoning trace (in text format) only and prompt it to categorize each sample based on how its emotion matches the context.
Figure~\ref{fig:errors} shows the result for each model and Figure~\ref{fig:classif_reason} provides with the prompt used.
While IndexTTS2 shows a large number of samples that convey the appropriate emotion, it also shows the largest number of samples with the wrong or opposite emotion.
Providing contextual information to IndexTTS2 largely reduces the ratio of wrong/opposite emotions.
Similarly, SparkTTS trained from scratch with contextual information that exhibits the largest ratio of samples with appropriate emotions across all variants of SparkTTS.

Overall, these results suggest that both models and human samples seem to insufficiently depict the appropriate emotions for a given quotation, showcasing challenges in faithfully impersonating characters during book reading.

Indeed, LibriVox recordings are recorded by amateur audiobook narrators, thus not all utterances are recorded with the same effort to deliver a perfect reading, which is the reason why ContextMOS ground-truth scores are not considerably higher.
We believe that replicating our experiments with professional audiobook narrators samples would likely exhibit higher ContextMOS, drawing a clearer gap between human and synthesized quotations.

\section{Hardware Information}
\label{app:5}

All experiments done in this paper were conducted on a single compute node equipped with 4 NVIDIA H100 with 80GB of GPU RAM.

\begin{figure}[t!]
    \small
    \begin{framed}
You will be given a paragraph that includes reasoning traces of a previous audio analysis. This analysis was performed by a LLM that was tasked to judge wheter a speech sample of an audiobook utterance was appropriately delivered in terms of prosody and emotion, given the book context in which the utterance occurs.

    Given the reasoning trace and the assessed score on a scale from 1 to 5, where 1 indicates complete failure at delivering the appropriate emotion, and 5 indicates complete success, your goal is to **classify** the prediction in **4 error categories**:
    \newline
    
   **Error Categories:**

        1. Success: the speech sample successfully delivered the right prosody and emotion.
        
        2. Opposite: the speech sample delivered natural prosody and emotion, but it was the opposite emotion/prosody that was expected given the context.
        
        3. Insufficient: the speech sample delivered insufficient prosody and emotion given the context.
        
        4. Wrong: the speech sample delivered natural prosody and emotion, but it was not the one expected given the context.
        \newline

    Now, you will be provided the reasoning trace, that include the prediction reasoning in natural language, and the integer score in a 1-5 scale.
    \newline

    **Reasoning Trace:**
    
    \{\{\{reasoning\_trace\}\}\}
    \newline

    **output\_format**
    
        You will output a json dictionary as follows:
        
        \{
        
        "error\_reasoning": str = Reasoning chain to explain why the prediction should be either one of the **Error Categories**.
        
        "prediction": int = Your prediction between 1 and 4 based on the **Error Categories**.
        
        \}
        \newline
        
        - Note: Ensure the json structure is followed and the json output **MUST** be parsable without errors.(For example, escape the quotes wherever you add them inside a field of the json, all brackets and braces should be correctly paired.)
    \end{framed}
        \caption{Example prompt used with Gemini-2.5 Flash for predicting error categories.}
\label{fig:classif_reason}
\end{figure}

\begin{table}[b!]
    \centering
    \small
    \begin{tabular}{cccc}
    \toprule
Admit & Announce & Argue & Assure \\
Babble & Bark & Bawl & Beg \\
Bellow & Bemoan & Blabber & Bleat \\
Bluster & Boast & Brag & Breathe \\
Cackle & Chant & Cheer & Chirp \\
Chirrup & Cluck & Complain & Confide \\
Cough & Drawl & Exclaim & Falter \\
Fuss & Giggle & Groan & Grumble \\
Growl & Grunt & Hiss & Holler \\
Hoot & Howl & Hum & Implore \\
Jabber & Jibber & Laugh & Moan \\
Mouth & Mumble & Murmur & Mutter \\
Nag & Pant & Pester & Prattle \\
Pronounce & Ramble & Rebuff & Retort \\
Roar & Sass & Scream & Screech \\
Shout & Shriek & Sing & Sigh \\
Snap & Snarl & Snicker & Sniff \\
Snigger & Snivel & Sob & Spit \\
Sputter & Squeak & Squeal & Stutter \\
Taunt & Tease & Trill & Wail \\
Weep & Whimper & Whine & Whisper \\
Whistle & Yell & Yelp & Hesitate \\
Pause \\
\bottomrule
    \end{tabular}
    \caption{List of speech verbs used for building the filtered subset $\mathbf{Q}_f$.}
    \label{tab:speech_verbs}
\end{table}

% \newpage 

\begin{figure*}
\small
\begin{framed}
You are an expert in linguistic. You like to read books and excel at analyzing dialogues in literature.
\newline

Given a small narrative passage, where each quote content is masked, your role is to extract speech verbs, adverbs, adjectives and nouns that indicate how a target quotation is being uttered. You will be given a target quotation marked with [TARGET] that occur in the passage. You need to extract speech verbs, adverbs, adjectives and nouns that follow these criteria:

- It must be either a speech-verb, an adverb, a noun or an adjective.

- It must be one word only.

- It must be a speech descriptor of the target quotation.

- If an adverb, it must be a descriptor of one of the speech-verbs describing the target quotation.

- If a verb, \textbf{ensure that it is a speech-verb and not a verb describing anything else than speech.}

Note that multiple speech-verbs can be found and that target quotations can have no associated speech-verbs.
\newline

Return a dictionary where keys are the words extracted in the final step and the values is another dictionary with keys 'id' for the paragraph id of the word, 'type' for the word type (verb, adverb, adjective, noun) and 'confidence' an integer between 0 and 10 measuring how confident you are in your prediction, 0 being not confident at all and 10 being sure you are right.

Before creating the dictionary, make sure again that all the criteria above are respected.
\textbf{Only generate this dictionary and nothing else. Return an empty dictionary if no words were found.}
\newline

Passage:

[0] Ella handed the notebook to Jay, eyes uncertain.

[1] Jay flipped through the sketches, pausing at one. [QUOTE\_1] She nodded.

[2] [TARGET] whispered Ella slowly.
\newline

Target quotation: [TARGET]
\newline

Answer:

\{
    "whispered": \{
        "id": "2",
        "type": "verb",
        "confidence": 10
    \},
    "slowly": \{
        "id": "2",
        "type": "adverb",
        "confidence": 10
    \}
\}
\newline

Passage:

[0] She went on, half laughing

[1] [TARGET] Then we went to the park, and he said [QUOTE\_1]
\newline

Target quotation: [TARGET]
\newline

Answer:

\{
    "went": \{
        "id": "0",
        "type": "verb",
        "confidence": 9
    \},
    "laughing": \{
        "id": "0",
        "type": "verb",
        "confidence": 9
    \}
\}
\newline

\dots
\newline

Passage:

[0] I said I had got it on the boat. So then she started for the house, leading me by the hand, and the children tagging after. When we got there she set me down in a split-bottomed chair, and set herself down on a little low stool in front of me, holding both of my hands, and says:

[1] [TARGET]

[2] [QUOTE\_1]
\newline

Target quotation: [TARGET]
\newline

Answer: 
\end{framed}
        \caption{Example prompt used with \texttt{Phi-4} (with self-reported confidence) to extract narrative information.}
\label{fig:prompt}
\end{figure*}

\begin{figure*}
    \scriptsize
    \begin{framed}
Your goal is to judge two audio samples that deliver the same text and analyze if a speech sample seems better than the other one on a particular **evaluation dimension** and determine the winner based on the scoring criterion.
    You will rate each sample a score between 0 and 3 based on how well the speech corresponding to the target quotation within a particular text called **contextual\_text** is delivered, then do their comparative analysis and provide your final judgement.
    A sample should sound realistic and human-like, and should capture the specific nuances of the text.
    \newline
    
    You will be provided with the **contextual\_text** which is the full text containing the quotation delivered in the speech sample and other side information. 
    the **text\_category** and the **evaluation\_criterion** corresponding to the **text\_category**, in which you will be made aware of the **evaluation dimension** you will focus on, and the **scoring criteria** you will use to score the samples.
    You will also be provided with the **output\_format**, which dictates the format of the output you need to follow as a judger.
    Finally, you will first be provided with the speech sample 1 **sample\_1** and then speech sample 2 **sample\_2**.
    \newline
    
    **contextual\_text**

    \{\{\{context\}\}\}
    \newline
    
    **text\_category**
    
    Emotions

    **evaluation\_criterion**
    
        **Evaluation Dimension:** 
        
            - In this category, we want the expression of natural emotions and contextual voice acting, using variations in pitch, loudness, rhythm, etc.
            
            - The sample contain speech delivered from a quotation present in the **contextual\_text** and identified with <quote\_start> and <quote\_end> markers. Delivering a good quality speech means showing natural and strong expressiveness for the quoted dialogue, which aligns with the contextual information.
        \newline
        
        **Example:**
        \newline

        \{\{\{Examples\}\}\}
        \newline
        
        **Rating Scale:**
        
        1: Fails to express emotions or contextual voice acting (whispering etc..).
        
        2: The rendered emotions or contextual voice acting are not very natural.
        
        3: Synthesises the quotation with natural emotions or contextual voice acting.

        **Note:**
            - The **contextual\_text** will not explicitly state the emotion for the quotation, you have to infer that from the context.
            - Samples demonstrating exaggerated expressiveness should not be rewarded more **UNLESS** the expressiveness features are relevant to the delivery of appropriated emotion and/or voice acting.
        \newline
        
        **Reasoning guidelines:**
        \newline
        \{\{\{reasoning\_guidelines\}\}\}
        \newline
  
    NOTE: If the speech is very poor and does not synthesise the text correctly, you will provide a score of 0.
    GLOBAL CONSIDERATIONS(**VERY IMPORTANT FOR COMPARISON**): 
        - It is imperative to compare the two samples **ONLY** on the basis of the **evaluation\_dimension**, that means, you **WILL NOT** let the following types of **BIASES** affect your judgement: 
        \newline

        \{\{\{bias\_mitigation\}\}\}
        \newline

    **output\_format**
    \newline
    
    You will output a json dictionary as follows:
    
    \{
    
    "reasoning\_sample\_1": str = ...
    
    "reasoning\_sample\_2": str = ...
    
    "comparison": str = ...
    
    "score\_1": int = ...
    
    "score\_2": int = ...
    
    "winner": int = ...
    
    \}

    Now you will be provided with the speech from sample 1, please analyze it carefully.
    \newline
    
    **sample\_1**
    
    \{\{\{audio\_1\}\}\}
    \newline

Now you will be provided with the speech from sample 2, please analyze it carefully. After that provide the judgment following the **output\_format** ensuring parsability.
\newline

    **sample\_2**
    
    \{\{\{audio\_2\}\}\}

    \end{framed}
        \caption{Example prompt used with Gemini-2.5 Pro for Win-Rate. Provided audios are raw wav files.}
\label{fig:win_rate}
\end{figure*}

\begin{figure*}
    \small
    \begin{framed}
Your goal is judge a single audio sample that deliver a specific quotation text. You will be provided with the **contextual\_text** which is the full text containing the quotation delivered in the speech sample and other side information.
    The target quotation that is spoken in the speech sample will be identified with <quote\_start> and <quote\_end> markers.
    Given the **contextual\_text** and the target quotation, you will rate the sample on a score between 0 and 5, based on how well the corresponding speech sample good quality speech means shows natural and strong expressiveness for the quoted dialogue, which aligns with the contextual information.
    In particular you will focus on aspects of emotions and voice acting and ignore any other factors.
    You will also be provided with the **output\_format**, which dictates the format of the output you need to follow as a judger.
    Finally, you will be provided with the speech sample **sample**.
    \newline
    
    **contextual\_text**

    \{\{\{context\}\}\}
    \newline

    **evaluation\_criterion**
    
- We want to evaluate the expression of natural emotions and contextual voice acting, using variations in pitch, loudness, rhythm, etc.
        \newline
        
        **Example:**
        \newline

        \{\{\{examples\}\}\}
        \newline
        
        **Rating Scale:**
        
1: Fails completely to express emotions or contextual voice acting (whispering etc..).

        2: Delivers an emotion or contextual voice acting that does not match the contextual information.
        
        3: The rendered emotions or contextual voice acting are not very natural.
        
        4: The rendered emotions or contextual voice acting is matching the contextual information but is not strong enough.
        
        5: The delivered speech contains very natural emotions or contextual voice acting that perfectly align with the contextual information.
        \newline
        
NOTE: If the speech is very poor and does not match the text correctly, you will provide a score of 0.
    GLOBAL CONSIDERATIONS(**VERY IMPORTANT FOR COMPARISON**):
    
        - It is imperative to judge the sample **ONLY** on the basis of the **evaluation\_criterion**, that means, you **WILL NOT** let the following types of **BIASES** affect your judgement: 
        
            - The acoustical quality of the audio, background noise or clarity.
            
            - The gender and timbre features of the speaker.
            - Any other factors that are not related to the **evaluation\_criterion**.
            
            - Samples demonstrating exaggerated expressiveness should not be rewarded more **UNLESS** those features are relevant to the **evaluation\_criterion**
        \newline
        
        **Reasoning guidelines:**
        \newline
        \{\{\{reasoning\_guidelines\}\}\}
        \newline
  
    NOTE: If the speech is very poor and does not synthesise the text correctly, you will provide a score of 0.
    GLOBAL CONSIDERATIONS(**VERY IMPORTANT FOR COMPARISON**): 
        - It is imperative to compare the two samples **ONLY** on the basis of the **evaluation\_dimension**, that means, you **WILL NOT** let the following types of **BIASES** affect your judgement: 
        \newline

        \{\{\{bias\_mitigation\}\}\}
        \newline

    **output\_format**
    \newline
    
    You will output a json dictionary as follows:

    \{
    
    "reasoning": str = Reasoning chain based on the **Reasoning guidelines:**.
    
    "score": int = Your score for the speech sample between 0 and 5, based on the **evaluation\_criterion** and what you have mentioned in "reasoning".
    
    \}

    Now you will be provided with the speech sample, please analyze it carefully.
    \newline
    
    **sample**
    
    \{\{\{audio\}\}\}
    \end{framed}
        \caption{Example prompt used with Gemini-2.5 Pro for ContextMOS. Provided audios are raw wav files.}
\label{fig:ctxtmos}
\end{figure*}

% \begin{figure*}
% \centering
%     \includegraphics[width=.9\linewidth]{cmos.png}
%     \caption{Screenshot of the platform used for the CMOS experiment.}
%     \label{fig:cmos_platform}
% \end{figure*}

% \begin{figure*}
% \centering
%     \includegraphics[width=.9\linewidth]{smos.png}
%     \caption{Screenshot of the platform used for the MOS experiment.}
%     \label{fig:mos_platform}
% \end{figure*}
% \newpage 

\end{document}